\documentclass[pdflatex,sn-mathphys-num]{sn-jnl}
\usepackage{graphicx}%
\usepackage{multirow}%
\usepackage{amsmath,amssymb,amsfonts}%
\usepackage{amsthm}%
\usepackage{mathrsfs}%
\usepackage[title]{appendix}%
\usepackage{xcolor}%
\usepackage{textcomp}%
\usepackage{manyfoot}%
\usepackage{booktabs}%
\usepackage{algorithm}%
\usepackage{algorithmicx}%
\usepackage{algpseudocode}%
\usepackage{listings}%
\usepackage{graphicx}
\usepackage{float} 
\usepackage[version=4]{mhchem}
\theoremstyle{thmstyleone}%
\begin{document}
\title[Sparing of DNA irradiated with UHDR under Physiological Oxygen and Salt]{Sparing of DNA irradiated with Ultra-High Dose-Rates under Physiological Oxygen and Salt conditions}
\author*[1]{\fnm{Marc Benjamin} \sur{Hahn}}\email{marc-benjamin.hahn@fu-berlin.de}
\author[2]{\fnm{Sepideh} \sur{Aminzadeh-Gohari}}
\author[2]{\fnm{Anna} \sur{Grebinyk}}
\author[2]{\fnm{Matthias} \sur{Gross}}
\author[2]{\fnm{Andreas} \sur{Hoffmann}}
\author[2]{\fnm{Xiangkun} \sur{Li}}
\author[2]{\fnm{Anne} \sur{Oppelt}}
\author[2]{\fnm{Chris} \sur{Richard}}
\author[2]{\fnm{Felix} \sur{Riemer}}
\author[2]{\fnm{Frank} \sur{Stephan}}
\author[2]{\fnm{Elif} \sur{Tarakci}}
\author[2]{\fnm{Daniel} \sur{Villani}}
\affil*[1]{\orgdiv{Institut f\"ur Chemie}, \orgname{Universit\"at Potsdam}, \orgaddress{\street{Karl-Liebknecht-Str. 24-25}, \postcode{14476}, \city{Potsdam},  \country{Germany}}}
\affil[2]{\orgdiv{Deutsches Elektronen-Synchrotron}, \orgname{DESY}, \orgaddress{\street{Platanenallee 6}, \postcode{15738}, \city{Zeuthen}, \country{Germany}}}
\abstract{
Cancer treatment with radiotherapy aims to kill tumor cells while sparing healthy tissue. Therefore, the experimentally observed sparing of healthy tissue by the FLASH effect during irradiations with ultra-high dose rates (UHDR) enables clinicians to extend the therapeutic window and enhance treatment efficiency.
However, the underlying radiobiological and chemical mechanisms are far from being understood. 
DNA is one of the main molecular targets for radiation therapy. Ionizing radiation damage to DNA in water depends strongly on salt, pH, buffer and oxygen content of the solvent.
Here we present a study of plasmid DNA pUC19, irradiated with 18\,MeV electrons at conventional/low dose rates (LDR) and UHDR under tightly controlled ambient, as well as physiological oxygen conditions in phosphate-buffered saline (PBS) at pH\,7.4.
For the first time a sparing effect of DNA strand-break induction between UHDR ($>$10\textsuperscript{6}\,Gy/s) and LDR ($<$0.1\,Gy/s) irradiated plasmid DNA under physiological oxygen, salt and pH conditions is observed for total doses above 10\,Gy. Under physiological oxygen (physoxia, $\approx$5\,\%O\textsubscript{2}, 40\,mmHg), more single (SSB) and double strand-breaks (DSB) are observed when exposed to LDR, than to UHDR. This behaviour is absent for ambient oxygen conditions (normoxia, $\approx$21\,\%O\textsubscript{2}, 150-160\,mmHg).\\
The experiments are accompanied by Geant4/TOPAS-nBio based particle-scattering and chemical Monte-Carlo simulations to obtain detailed information about the yields of radicals and reactive oxygen species (ROS).
Hereby, an extended set of chemical reactions was considered, which improved upon the discrepancy between experiment and simulations of previous works, and allowed to predict DR dependent g-values of hydrogen peroxide (\ce{H2O2}).
To explain the observed DNA sparing effect under FLASH conditions at physoxia, the following model was proposed:
The interplay of \ce{O2} with the \ce{^.OH} induced hydrogen abstraction at the sugar-phosphate backbone, and the conversion of DNA base-damage to SSB, under consideration of the dose-rate dependent \ce{H3O^+} yield \textit{via} beta elimination processes is accounted for, to explain the observed behavior.
}
\maketitle
\section{Introduction}
\label{sec:intro}
The discovery of the so called FLASH effect opened up new opportunities to optimize the therapeutic window and thus make cancer treatment with radiation more efficient.
The FLASH effect is based on a favourable change of the differential radiation response between healthy and cancerous tissue, when increasing the dose rate (DR) during treatment from conventional values (LDR, below 2\,Gy/s) to ultra-high DR (UHDR, above 40\,Gy/s).
The healthy tissue sparing and relative increase of the sensitivity of different cancer-cell lines is currently being investigated in many settings, on the cellular level and in animal models. \cite{favaudonultrahigh2014,vozeninadvantage2019,fouilladeflash2020,friedlradiobiology2022,vozeninclinical2022,limolireinventing2023}
However, the underlying mechanisms remain unclear.
Various hypotheses on different levels were proposed, such as changes in the underlying radiation chemistry and radiation damage to biomolecular targets, influence of the repair capacities of radiation damage of UHDR, effects on the tumor microenvironment, immunosuppression and immune response, including combinations thereof.\cite{zhoumechanisms2020,friedlradiobiology2022,hahnaccessing2023}\\
From the point of view of physics and physical-chemistry, a high-precision control of temporal and spatial properties of the particle-beams used for irradiation, and a tightly controlled chemical environment are necessary, to test the various theories, and in consequence to move the field forward.\\
Since, the dose rate differences between LDR and UHDR span many orders of magnitude, especially UHDR irradiations are often performed with bunches or pulses in the sub-microsecond timescale.
With respect to the "chemistry related" hypotheses brought forward to explain the FLASH effect, many assume a modification of the oxygen related radiation chemistry and a change in the yield of reactive oxygen species (ROS), which damage biomolecules and can lead to cell killing.
Since the oxygen levels between healthy and tumor tissue can differ greatly, the interplay between DR and oxygen content in a cell may provide a ``chemistry based'' explanation for the observed FLASH effect.\\
To study such processes, the time scales of particle-matter interaction, in the sub-femtosecond timescale, the variation in  temporal and spatial homogeneity of the radiation bunches and bunch-trains in ranges of pico- to microseconds, as well as the radiation chemistry, diffusion processes and possible non-equilibrium dynamics of the involved reaction partners up to the range of microseconds, need to be considered.
To do so, a combination of experimental and simulational tools can provide insight on length and time scales which are difficult to access by the respective isolated methods alone.\cite{hahnmeasurements2017,hahnaccessing2023} \\
In the present work we study DNA irradiated with MeV electrons at conventional LDR and UHDR under ambient (normoxic, 21\,\% \ce{O2}) and the physoxic (5-6\,\% \ce{O2}) oxygen conditions in physiological pH and salinity from the experimental and simulational point of view.
DNA as the carrier of genetic information was chosen as an endpoint, since it plays a central role for transcription, mutation and apoptosis in living organisms - therefore DNA is the most important target for radiation damage on the molecular level.\cite{vonsonntagfreeradicalinduced2006,hahnaccessing2023}
That is the reason why so far most of the studies involving isolated biomolecules to compare the effects of LDR and UHDR in \textit{in-vitro}  were performed on DNA.
These studies applied various types of radiation sources such as protons,\cite{ohsawadna2022,wanstallquantification2023,konishiinduction2023}, electrons,\cite{wanstallquantification2023,smallevaluating2021,kacemunderstanding2022,perstinquantifying2022} and photons.\cite{sforzaeffect2024}
In these studies mostly DNA strand-breaks, and sometimes base damage were analyzed in dependence of the dose and DR.
Most of the studies incorporated Tris(hydroxymethyl)aminomethan (Tris) to control the pH in the solvent, whereby some added additional radical scavengers, besides Tris, at different concentrations to test hypotheses about the underlying radiation chemistry.\cite{wanstallquantification2023,sforzaeffect2024}
However, so far all these studies lack the presence of physiological salt concentrations.
This is a major drawback which is overcome by the present work, since the presence of salt is  crucial for DNA stability, realistic radiation chemistry and the DNA strand-break yield.\cite{vonsonntagfreeradicalinduced2006,hahndna2017}
Furthermore, with the exception of the work by Rezaee, Adhikary and coworkers, all studies were performed at ambient oxygen conditions.\cite{sforzaeffect2024}
These conditions do not represent the physoxic oxygen conditions of healthy cells, in which the sparing effect of of healthy tissue exposed to UHDR (the FLASH effect) is observed.
Therefore providing a tightly controlled, physiological oxygen environment is imperative for obtaining meaningful insights into the radiation chemical mechanisms underlying the FLASH effect.\\
To guarantee us to experimentally control the oxygen in solution within our experimental setup during the measurements, extensive measurements on sealing of sample tubing were performed to establish a safe duration within which these conditions can be controlled within necessary accuracy.\
The irradiation experiments themself were performed at at the Photo Injector Test Facility at DESY in Zeuthen (PITZ), where plasmid DNA pUC19 was irradiated with 18\,MeV electrons at LDR and UHDR in phosphate-buffered saline (PBS) buffer, providing physiological pH and salt concentrations, at atmospheric and physiological, tightly controlled oxygen conditions. 
The DNA radiation damage was analyzed in terms of DNA single (SSB) and double strand-breaks (DSB).
The electron beam structure within and dose deposit in the applied experimental setup was studied by Monte-Carlo scattering simulations (MCS) and cross validated against the measured doses during the experiments.
\section{Materials and Methods}
\label{sec:methods}
\subsection*{Plasmid DNA samples}
\label{sec:plasmiddna}
Double-stranded DNA (dsDNA), pUC19 plasmids with 2686\,base pairs (bp), at a concentration of 500\,ng\,$\mu$L\textsuperscript{-1} with a high degree ($>$90\,\%) of undamaged, supercoiled (SC) plasmids in 1$\times$PBS was obtained from Plasmidfactory (Germany).
The PBS buffer guarantees a physiological pH of 7.4 and salt concentrations, consists of Millipore water with 137\,mmol L\textsuperscript{-1} NaCl, 2.7\,mmol L\textsuperscript{-1} KCl, 10\,mmol L\textsuperscript{-1} Na\textsubscript{2}HPO\textsubscript{4} and 1.8\,mmol\,L\textsuperscript{-1} KH\textsubscript{2}PO\textsubscript{4}.
All DNA samples were stored below -20\,\textdegree\,C between the experiments and transported on dry ice when moved between laboratories for irradiation and analysis.
\subsection*{Irradiations}
\begin{figure}[!htbp]
\centering
\includegraphics[width=0.47\textwidth]{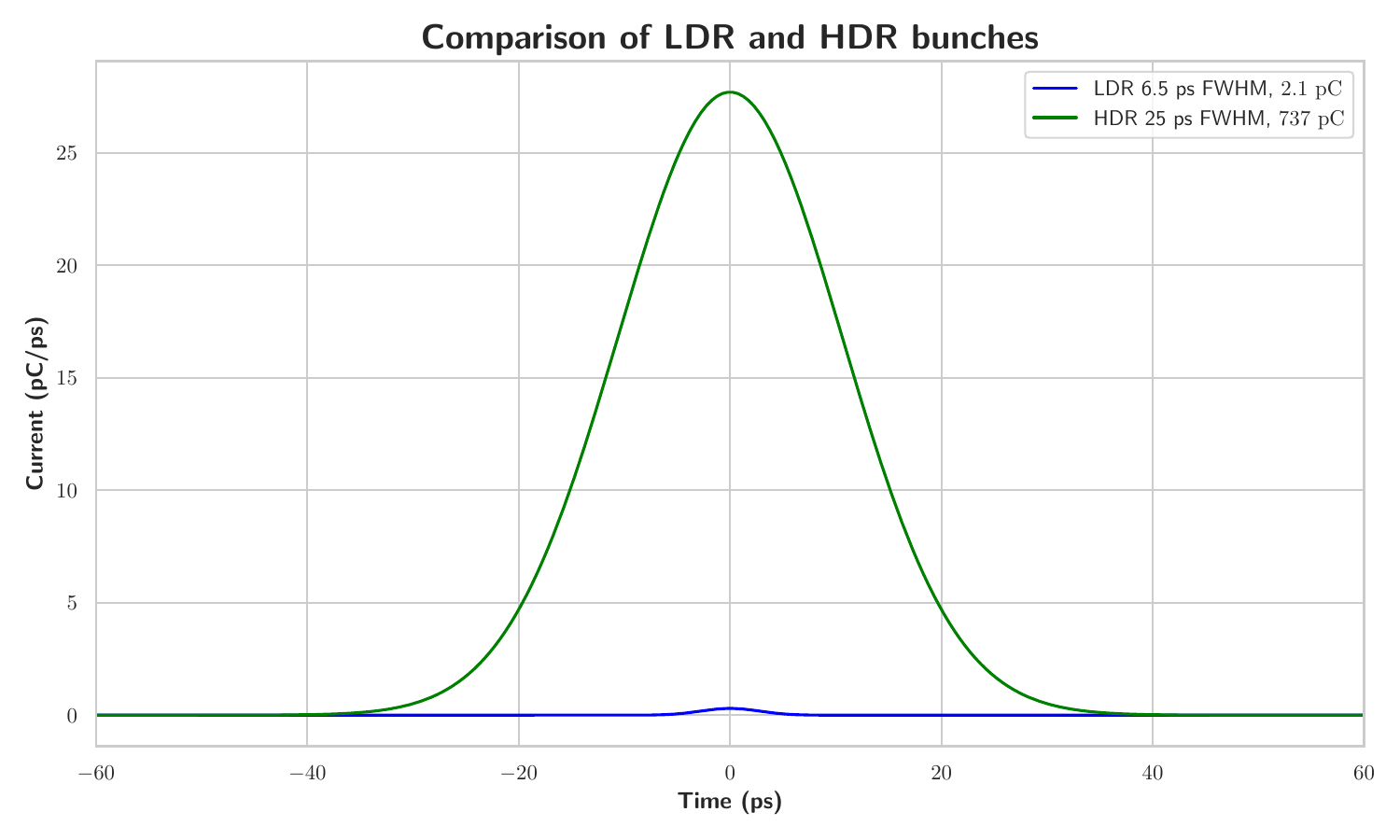}\includegraphics[width=0.53\textwidth]{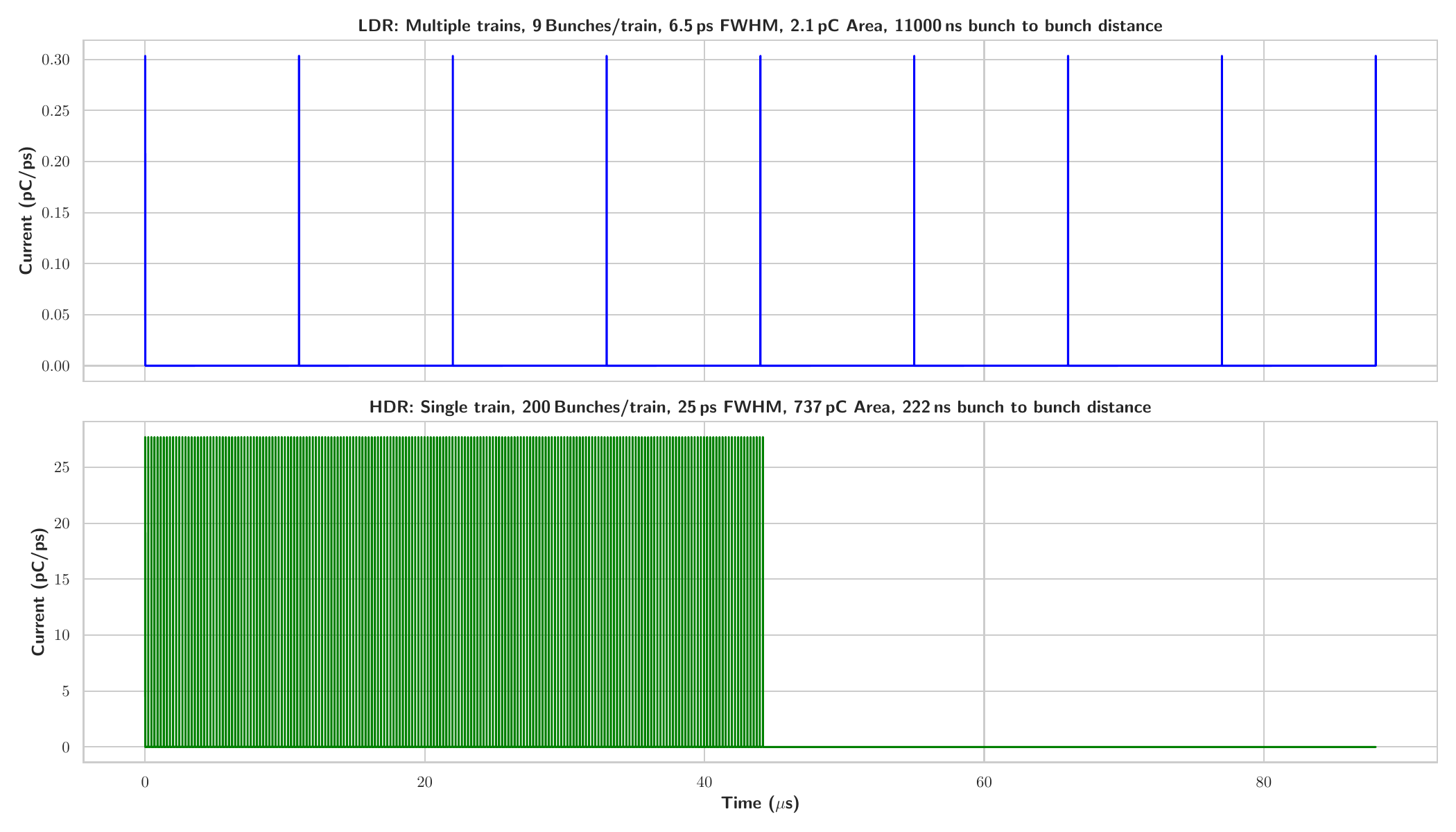}
\caption{\textbf{Electron bunches and trains:} Comparison of LDR (blue) and UHDR (green) bunch (left) and bunch-train (right) structure in time. Note the differences in y axis scale.}
\label{fig:bunchtrainstructure}
\end{figure}

\begin{figure}[!htbp]
\centering
\includegraphics[width=0.8\textwidth]{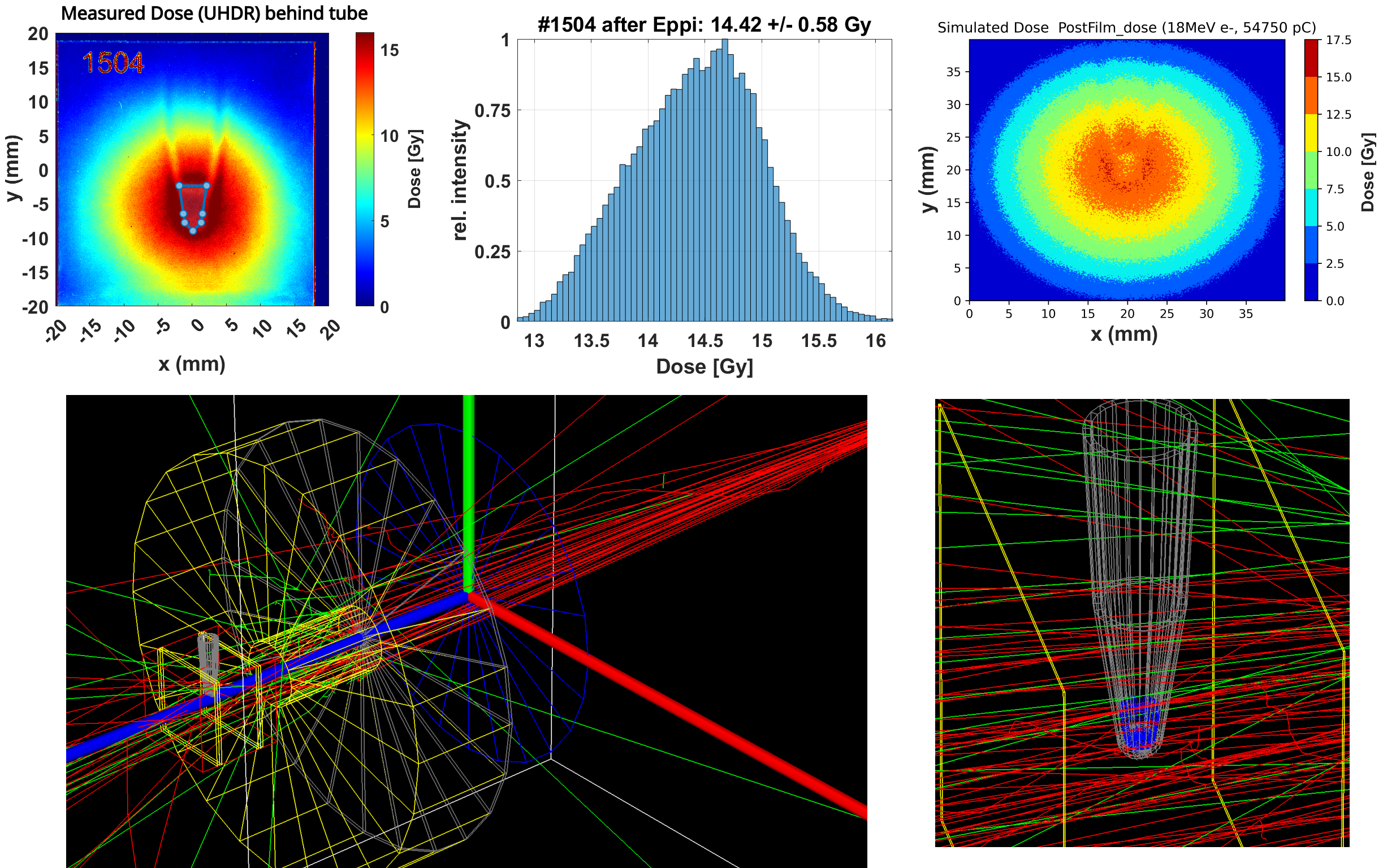}\\
\includegraphics[width=0.8\textwidth]{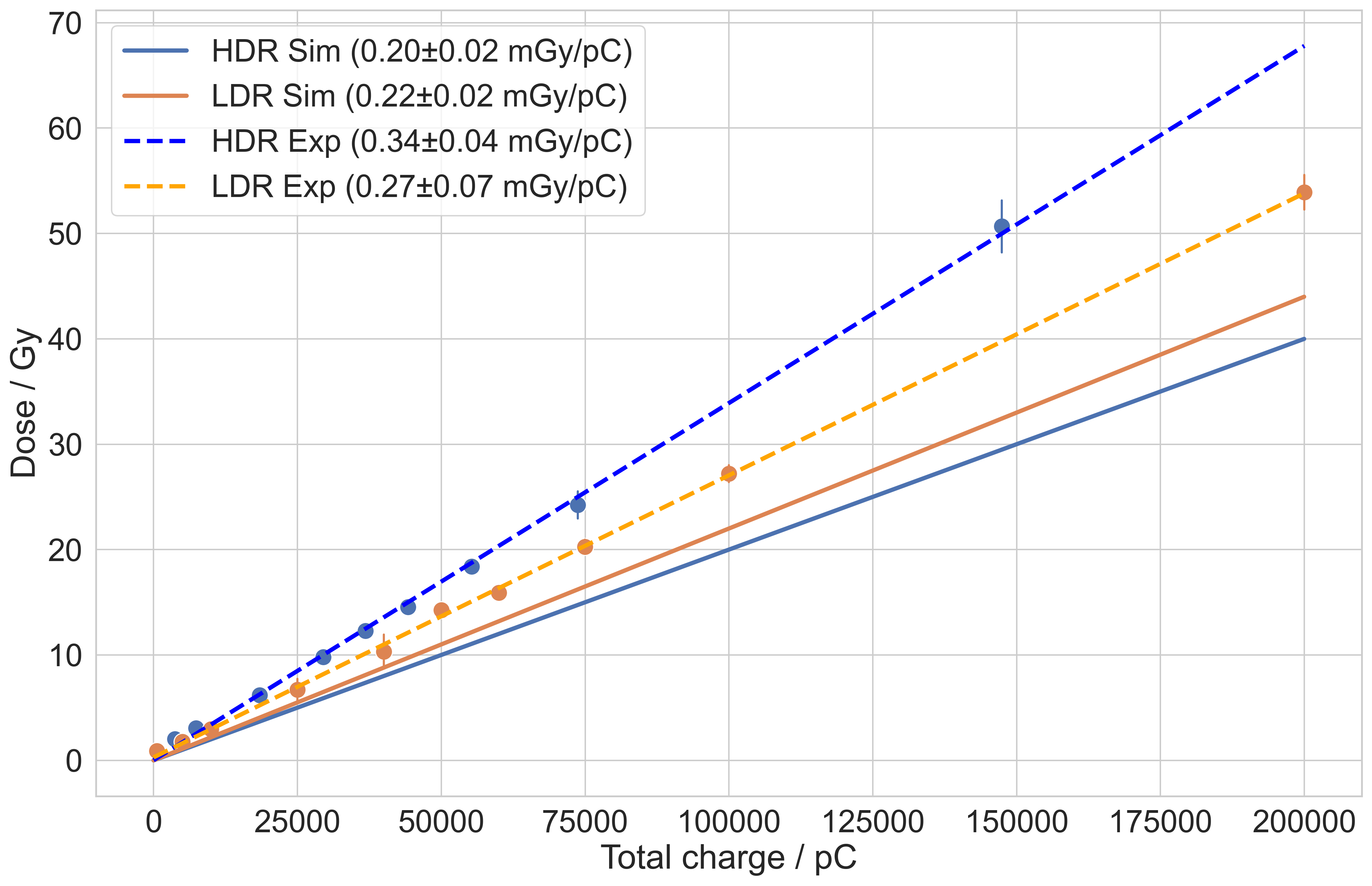}
\caption{\textbf{Experimental and simulated doses:} Comparison of measured and simulated dose: Top left: Exemplary spatial dose distribution measured on a Gafchromic film behind the sample tube with 0.5\,ml volume filled with 50\,$\mu$L sample (left). The region of interest for sample dose determination is marked by light blue dots and lines. Top Center: Histogram of the dose distribution within the measured target volume in the sample tube as indicated by the blue lines in the image on the left. Top right: Simulated dose distribution within the same Gafchromic film as obtained by Geant4/Topas particle scattering simulations for the same conditions as in the experiment.
Central image, left: Particle scattering simulations showing the simulated setup. Hereby electrons are shown in red, photons in light green. Sample tubes are grey, water blue, Gafchromic films before and behind the samples are shown in yellow. Before them, the scattering plate in blue and the lead shielding with a hole are shown in yellow. The image in the center right, shows a zoom to the sample tube and Gafchromic films. The electron beam propagated from right to left and for illustration purposes the number of electron simulated is strongly reduced with respect to the real case.
Bottom: Shown is the dose deposit within the water volume for LDR (orange) and UHDR (blue) conditions in experiment (circles and a linear regression as broken lines) and simulation (solid lines) in dependence of the total charge.}
\label{fig:filmdosimetry}
\end{figure}

Electron irradiation experiments were performed at the Photo Injector Test Facility (PITZ) at the Zeuthen branch of the \textit{Deutsches Elektronen-Synchrotron} (DESY) at the beamline of the \textit{FLASHlab@PITZ}, which was described previously in detail.\cite{stephanflashlab2022,lidemonstration2025}
The overall parameters of the irradiations for conventional (LDR) and ultra-high dose-rates (HDR) with 18\,MeV electrons are listed in Tab.\,\ref{tab:irradiation}.
Irradiations at doses between 0.9\,Gy-57\,Gy, conventional ($<0.1\,Gy/s$) and UHDR ($>10^6\,Gy/s$) mean dose rates, and ambient (normoxic, 150-160\,mmHg) as well physiological (physoxic, 38-42\,mmHg) oxygen content were performed on plasmid pUC19 DNA samples in physiological buffer (see Sec.\,\ref{sec:plasmiddna}).
The LDR irradiations were performed by applying multiple bunch trains with 10\,Hz repetition rate.
These trains consist out of well separated bunches (11\,$\mu$s) with low, 2.1\,pC per bunch to minimize instantaneous bunch and train dose-rate, are visualized in Fig.\,\ref{fig:bunchtrainstructure}.
In contrast, UHDR irradiations were performed to maximize the number of bunches in a single train in close distance, separated by 222\,ns, with up to 737\,pC per bunch, to achieve the maximal instantaneous bunch and train dose-rate.
Detailed beam parameter ranges are summarized in Tab.\,\ref{tab:irradiation} according to the extended scheme  proposed by Sch\"uler \textit{et al.}\cite{schulerultrahigh2022}
\begin{table}[!htbp]
\centering
\caption{Overview of the parameter ranges used for the irradiations, reported according to the scheme proposed by Sch\"ueler \textit{et al.}\cite{schulerultrahigh2022}, which was extended to include the parameters for bunch trains. All measured values are given with respect to the last significant digit if not given otherwise.}
\label{tab:irradiation}
\begin{tabular}{llll}
\toprule
Parameter & Unit & LDR & UHDR \\
\midrule
Total absorbed dose & Gy & 1.3-53.6 & 1.3-50.0 \\
Mean dose rate & Gy/s & 0.05 & $1.1\times10^{6}$ - $1.4\times10^{6}$ \\
Duration of exposure & s & 25.90 - $1.06\times10^{3}$ & $8.89\times10^{-7}$ - $4.42\times10^{-5}$ \\
Total charge & pC & $4.91\times10^{3}$ - $2.00\times10^{5}$ & $3.69\times10^{3}$ - $1.47\times10^{5}$ \\
Total electrons & - & $3.07\times10^{10}$ - $1.25\times10^{12}$ & $2.30\times10^{10}$ - $9.20\times10^{11}$ \\
Dose per train & Gy & $5.1\times10^{-3}$ & 1.3 - 50.0 \\
Train dose-rate & Gy/s & 57.6 & $1.1\times10^{6}$ - $1.4\times10^{6}$ \\
Train length & s & $8.80\times10^{-5}$ & $8.89\times10^{-7}$ - $4.42\times10^{-5}$ \\
Charge per train & pC & 18.90 & $3.69\times10^{3}$ - $1.47\times10^{5}$ \\
Number of trains & - & 260 - $1.06\times10^{4}$ & 1 \\
Train frequency & Hz & 10.00 & - \\
Dose per bunch & Gy & $5.6\times10^{-4}$ & 0.25 \\
Instant. (bunch) dose-rate & Gy/s & $8.7\times10^{7}$ & $1.0\times10^{10}$ \\
Bunch width (FWHM) & ps & 6.50& 25.0 \\
Beam size (rms) & mm & 1.1 - 1.4 & 3.7 - 4.9 \\
Charge per bunch & pC & 2.1$\pm$0.6 & 737$\pm$6 \\
Bunch separation in train & s & $1.10\times10^{-5}$ & $2.22\times10^{-7}$ \\
Bunch per train & - & 9 & 5 - 200 \\
Bunch frequency in train & Hz & $9.09\times10^{4}$ & $4.50\times10^{6}$ \\
Total energy & MeV & 17.97 & 18.03 \\
\bottomrule
\end{tabular}
\end{table}
Dosimetry during the measurements was performed based on Gafchromic EBT-XD films (Ashland),\cite{palmerevaluation2015} whereby one film is placed in front and one behind the sample, allowing to measure the entrance and exit dose distribution.
The calibration of the films was performed in a clinical beam at the Charit\'{e} hospital in the range from 0.1\,Gy up to 200\,Gy. The readout of the films was performed 24 hours post irradiation by using a calibrated \textit{Epson Expression 12000XL} scanner. The average value of all three color channels is used. To calculate the delivered dose at the sample location the calibration factor from in-tube measurements is used. Additional details can be found in a previous work.\cite{lidemonstration2025}
From these films, the spatial dose distribution within the irradiated volume can be determined as exemplarily shown in Fig.\,\ref{fig:filmdosimetry}. 
The differences between simulated and measured doses are most likely caused to a great degree from the deviation between the perfect ideal Gaussian form of the beam profile included in the simulation and the non-ideally and non-perfectly symmetrical distributed beam profiles throughout the experiments.
Furthermore, it was recently observed, that Gafchromic films can shown an overresponse under UHDR compared to LDR,\cite{delsartosystematic2025} which is in agreement with the measured difference in the dose deposit per pC for the experimental curves shown in Fig.\,\ref{fig:filmdosimetry}.
The inverse difference per pC during the simulations is a result of the spatially broader beam for UHDR compared to LDR, and therefore a number of lower hits within the ``target volume'' represented by the water in the tubes.
\subsection*{Oxygen control and stability}
\label{sec:oxygen}
\begin{figure}[!htbp]
\centering
\includegraphics[width=0.8\textwidth]{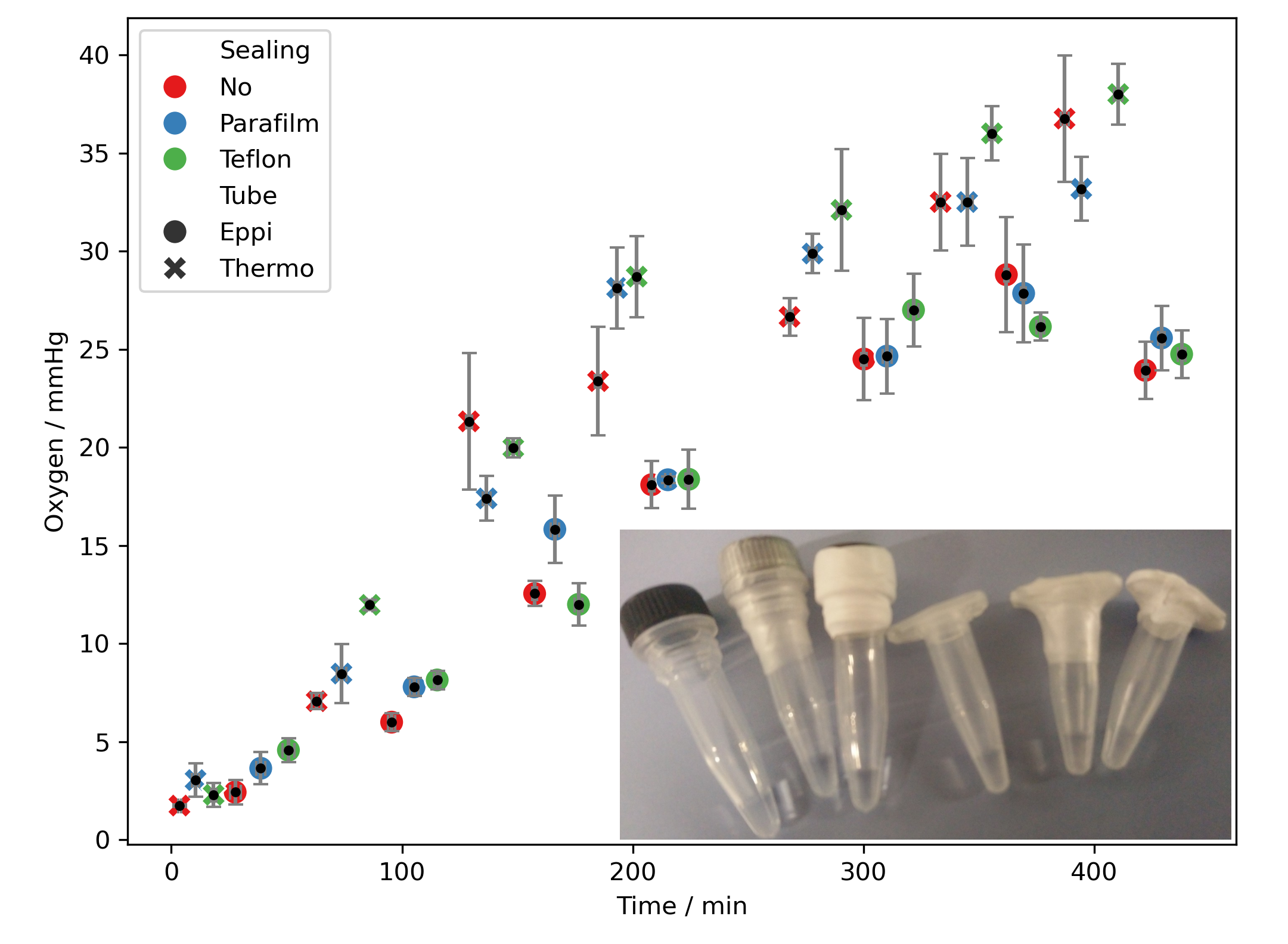}
\caption{\textbf{Oxygen stability:} Shown is the oxygen content over time for different tubes and sealing approaches for closed sample tubes stored under ambient oxygen at room temperature, measured directly with an Oxford Oxylite oxygen sensor after opening the respective tube.  The inset shows the Thermo and Eppendorf tubes without additional sealing and with Parafilm and Teflon sealing, from left to right respectively.}
\label{fig:Oxygensealing}
\end{figure}

For the irradiation experiments, samples with reduced (physiological) oxygen concentrations of 38-40\,mmHg ($\approx$65\,$\mu$M \ce{O2} physoxia, corresponding to $\approx5\,\%$ Oxygen in normal tissue) were prepared in a HypoxyLab oxygen chamber (Oxford Optronix, UK) by letting them equilibrate their dissolved oxygen content at room temperature under $\approx$95\,\,\%\,Rel.\,humidity.
The final concentration of 40\,mmHg oxygen content is achieved by automatically adjusted ratios of synthetic air, and N\textsubscript{2} laboratory gases (Nippon gasses purity of 99.999\,\%) within the oxygen chamber.
The oxygen content within the samples were checked with an Oxford Oxylite oxygen sensor (Oxford Optronix, UK) at technical replicates of the samples prepared in parallel to the samples to be irradiated under the same atmosphere.
The samples about which we refer in the following as ambient oxygen conditions where prepared under standard air (21\,\% \ce{O2}) which corresponds to around $\approx$270\,$\mu$M \ce{O2} and 150-160\,mmHg at room temperature in solution. The range given for the mercury equivalent values here, results from the used probe in the Oxylite sensor, which is optimized for measuring under low-oxygen conditions.\\
To test the optimal sealing of the sample containers against diffusion of oxygen for future experiments under low oxygen conditions, different types of sample tubes with various sealings were prepared at oxygen levels below $<$2\,mmHg (radiobiological hypoxia).\cite{mckeowndefining2014} Therefore common Eppendorf PCR and Thermo Fisher tubes, with either snap (Eppendorf) or screw caps with o-rings (Thermo) of 0.5\,mL volume were filled each with 50\,$\mu$L PBS buffer. 
These were either closed without sealing, or prepared with addition sealing of either Parafilm or Teflon, as shown exemplarily in the inset of Fig.\,\ref{fig:Oxygensealing}.
The Eppendorf tubes with snap-lock provided a more efficient sealing that the Thermo tubes with screw cap for durations of about three hours at oxygen conditions around 20\,mmHg oxygen in the solvent, while being exposed to ambient oxygen. The additional sealing by Teflon or Parafilm didn't show a substantial difference. 
Thus, to perform experiments at hypoxic conditions further improvement of the sealing and/or shortest possible exposure times are needed with a stringent combined monitoring of the oxygen levels during and after the experiments.
\subsection*{DNA damage analysis}
\begin{figure}[!htbp]
\centering
\includegraphics[width=0.8\textwidth]{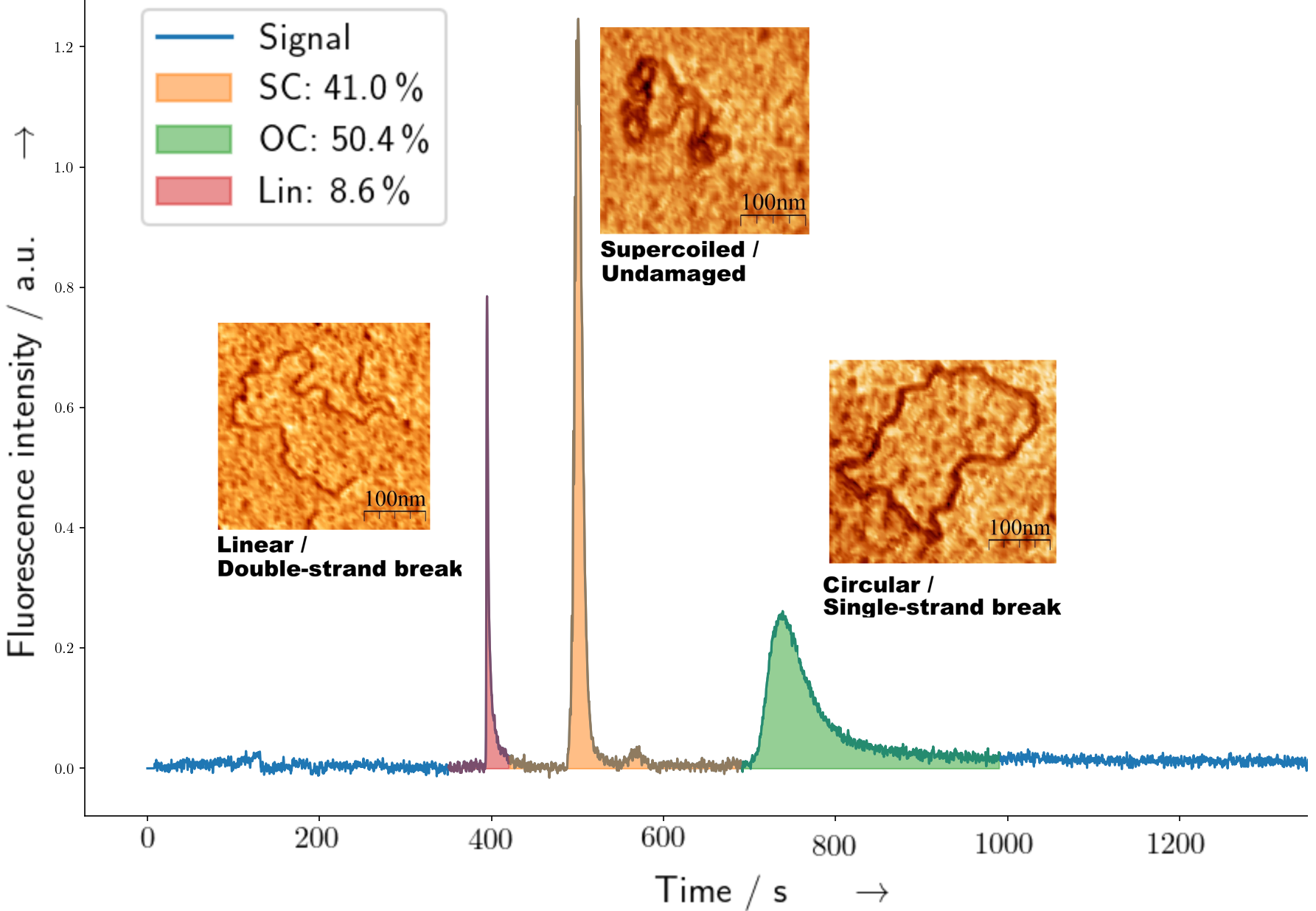}
\caption{\textbf{Plasmid conformations, damage and their position in an exemplary CGE measurement.} Shown is a CGE densiogram of plasmid pUC19 with 2686\,bp and three clearly visible isoforms. From left to right as linear (Lin: red, around 400\,s), supercoiled/covalently-closed circular (SC: yellow, around 500\,s) and open-circular (OC: green, around 750\,s) conformation. To visualize the microscopic structure of the different plasmid DNA isoforms, atomic-force microscopy (AFM) images are shown as insets next to the respective band in the densiogram. The images were taken and combined with permission from previous works.\cite{cordsmeierdna2022,hahnrapid2025}}
\label{fig:cgeplasmid}
\end{figure}

DNA isoforms and the related damage was analyzed by capillary gel electrophoresis (CGE).
The undamaged plasmids exists in a topological constrained form and it is supercoiled (SC). 
A single-strand break (SSB) at the DNA backbone leads to the relaxation of the SC plasmid to an open circular form (OC). 
When a double-strand break (DSB), consisting out of two SSBs at different sides of the dsDNA (assumed to be within a distance of around 10\,bp) occurs, the plasmid DNA relaxes even further to linear conformation (Lin). These different isoforms and types of damage, undamaged/SC, SSB/OC and DSB/Lin, are separated during a CGE measurement due to their different electrophoretic mobility, as shown in Fig.\,\ref{fig:cgeplasmid}. 
This makes CGE ideal to quantify the different isoforms and type of DNA damage in irradiated plasmid DNA samples.\cite{hahnrapid2025}\\
From the irradiated plasmid DNA and control samples with a concentration of 500\,ng/$\mu$L, aliquots for analysis with a concentration of  50\,ng/$\mu$L were prepared for the CGE analysis in QIaxcel dilution buffer.
Directly before the measurements the samples were briefly vortexed, and then centrifuged in a micro centrifuge.
These DNA samples were measured in a QIaxcel Advanced system (Qiagen, Germany) with a QIaxcel DNA Screening Kit, with a 40\,s injection time with a custom method (AM900-custom), providing an extended runtime.
This custom run profile uses the following run parameters: sample injection voltage of 2\,kV, an injection times of 40\,s, a separation voltage of 4\,kV, and a total separation time of 1440\,s.
The CGE data was exported to a CSV from the Qiagen Screen Gel software (1.6.0) and analysed after performing a background subtraction.
Afterwards, the regions assigned to the SC, OC and linear plasmid DNA forms were integrated, whereby the regions for integration of the linear, supercoiled and open-circular form are centered around 400\,s, 500\,s and 750\,s respectively. 
For the quantitative analysis only samples with an adequate SNR ratio and overall high signal intensity were included.
The integrated areas were normalized on the sum obtained for the SC, OC and Lin area within each dataset, details on the accuracy, validity and testing of the method were published previously.\cite{hahnrapid2025}
\subsection*{Particle scattering simulations}
\label{methods:mcs}
The particle scattering simulations were performed with the \textit{Geant4} 11.1.3 framework and the \textit{OpenTOPAS} 4.0.0 interface.\cite{agostinelligeant42003a,perltopas2012}
For the simulation of the beamline, the \textit{g4em-standard\_opt4} physics list was used, which performs simulations by a condensed history approach.
Additional options and processes were activated, namely \textit{Fluorescence},
\textit{Auger}, \textit{AugerCascade}, \textit{PIXE}, and \textit{DeexcitationIgnoreCut}, while \textit{ActivateMultipleScattering} was deactivated.
The water (volume 50\,$\mu$L) in the standard Eppendorf 0.5\,mL sample tube was simulated for standard conditions with the \textit{G4\_WATER} material and the sample tube with the material \textit{G4\_POLYPROPYLENE} and a wall thickness of 0.6\,mm.
Hereby 10\,pC of primary electrons were simulated, and the dose and dose distribution in water and the Gafchromic films were scored.
The composition of the active part of the Gafchromic film was simulated with 0.025\,mm thickness and a chemical composition of
 57.40\,\% Hydrogen, 0.6\,\% Lithium,  28.5\,\% Carbon, 0.4\,\% Nitrogen, 11.7\,\% Oxygen, 1.4\,\% Aluminum and a density of 1.35\,g/cm\textsuperscript{3}, while the polyethylene cover found on both sides were modeled with 0.125\,mm thickness each, and 36.4\,\% Hydrogen, 45.5\,\% Carbon, 18.2\,\% and Oxygen, in the respective relative fractions.\cite{palmerevaluation2015}
\subsection*{Simulation of the chemical stage}
\begin{figure}[!htbp]
\centering
\includegraphics[width=0.76\textwidth]{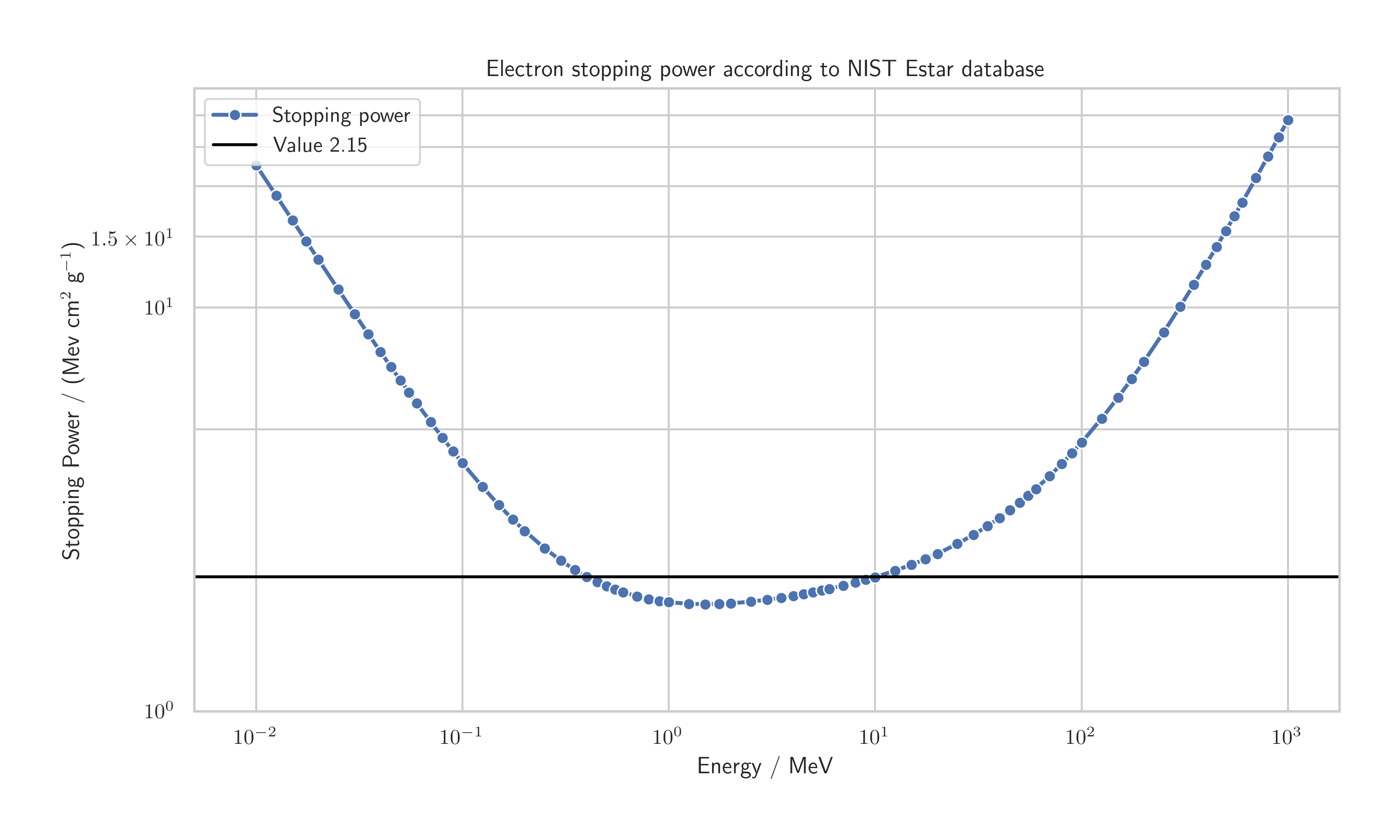}\\
\includegraphics[width=0.76\textwidth]{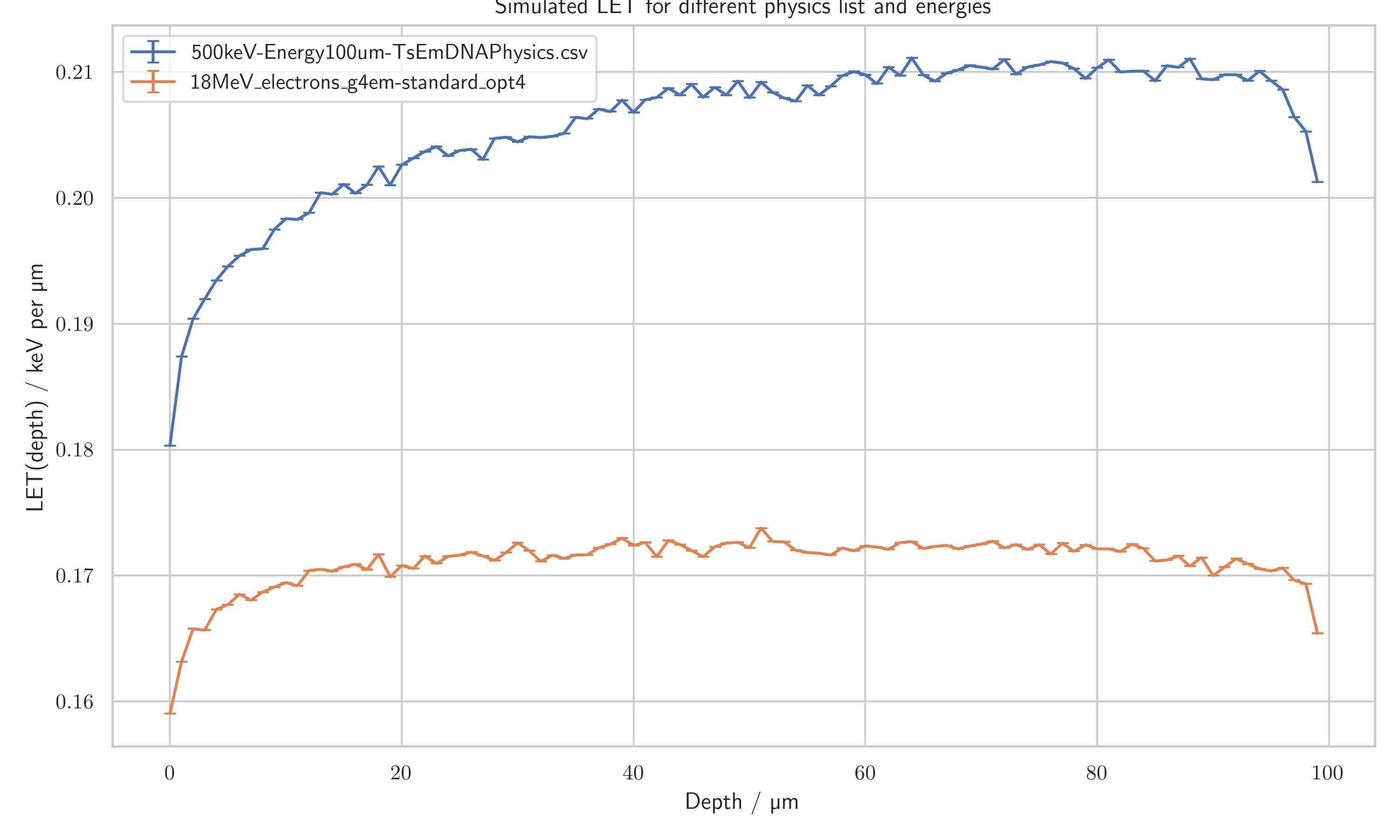}\\
\includegraphics[width=0.66\textwidth]{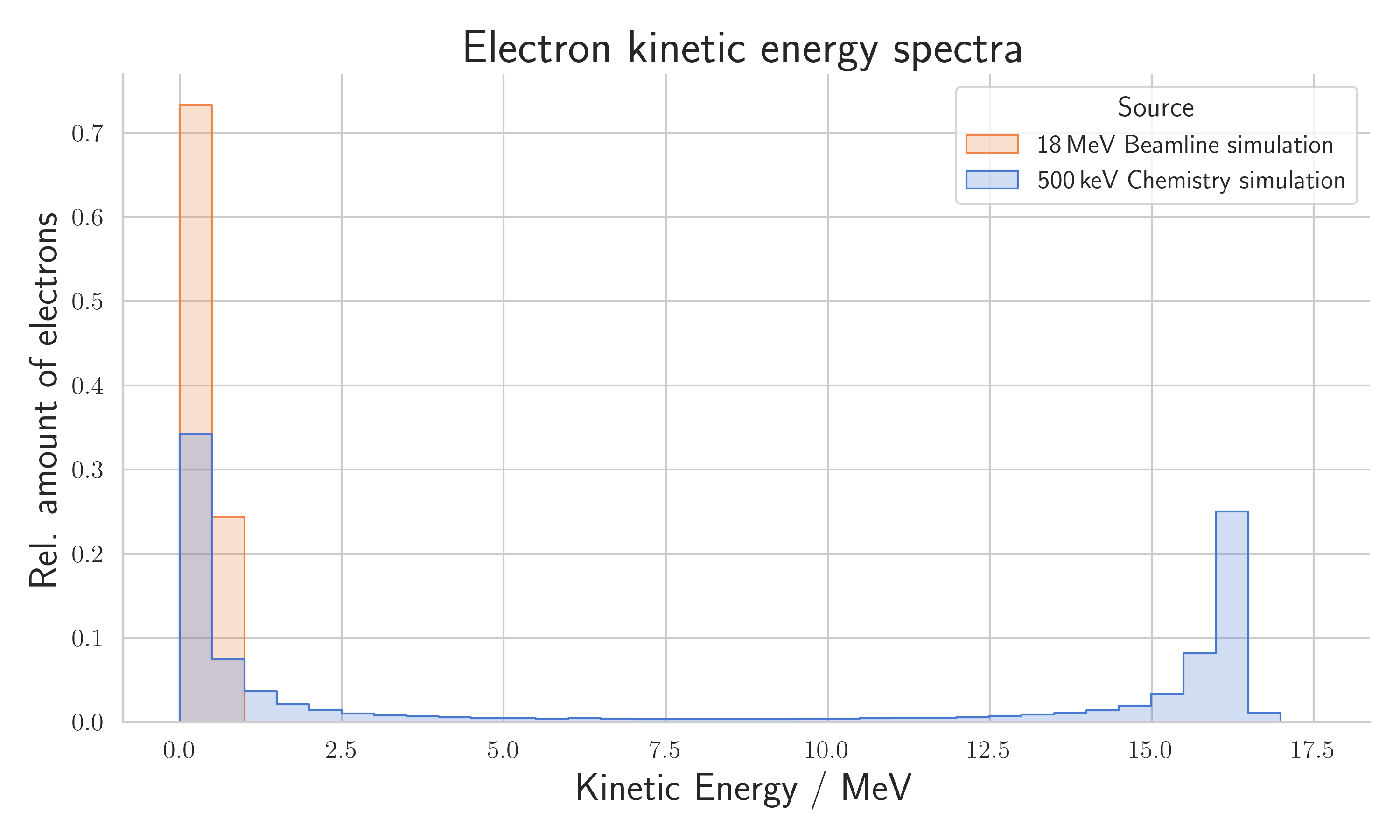}
\caption{\textbf{Properties of the electron beam:} Energy dependent stopping power of electrons from the \textit{NIST Elstar} database (top) and the related LET values (center) for different electron energies and simulation settings, and the respective kinetic energy spectra originating from particle scattering simulations (bottom). For simulation details see Sec.\,\ref{sec:methods} and the discussion.}
\label{fig:letelectrons}
\end{figure}

Simulations of the chemical stage were performed with \textit{Topas-nBio 4.0}\cite{schuemanntopasnbio2019} and the independent reaction time (IRT) modules, which allow for faster simulations of the  reaction processes compared to conventional step-by-step diffusion approaches.\cite{ramosmendezindependent2020,planteconsiderations2017}
To simulate the effects of different bunch charge, bunch width, bunch form the \textit{TsIRTInterPulse} scorer was used with varying parameters based on considerations from the literature.\cite{shinevaluation2019}
The \textit{TsEmDNAPhysics} and \textit{TsEmDNAChemistry} physics and chemistry lists were used, while the electron elastic scattering model \textit{ELSEPA} and solvated electron thermalization model \textit{Ritchie} was activated.
The time resolution for the chemical stage was set to 0.5\,ps, the test for contact reactions active, and the \textit{IRTProcedure} was set to pure, while the time step model was \textit{G4IRT}.
All other additional options and processes were the same as described in Sec.\,\ref{methods:mcs}
The chemical reactions, rate constants and scavenging capacity are summarized in Tab.\,\ref{tab:chemistry} and compiled from various literature sources. \cite{buxtoncritical1988,vonsonntagchemical1987,planteconsiderations2017}
The diffusion constant for the chemistry simulations were used as provided by \textit{Topas-nBio 4.0}.
All simulations were performed in a cubic target volume of 512\,$\mu$m\textsuperscript{3}.
During all simulations a 500\,keV \ce{e^-} beam entered the simulation volume randomly distributed over the X-Y plane and performed inelastic scattering interactions within the target volume filled with \textit{G4\_WATER}.
The electron energy of 500\,keV was chosen, since the more accurate scattering track structure models for water, provided by  \textit{Geant4-DNA} are applied for electrons with energies below 1\,MeV. 
These 500\,keV electrons have a similar LET (Fig.\,\ref{fig:letelectrons} top) as 18\,MeV electrons, when comparing the tabulated LET values of 18\,MeV electrons from the NIST Elstar database, with the LET values of the simulated 500\,keV ($\approx0.21\,keV/\mu m$) in the chemistry simulations performed with the step-by-step physics \textit{TsEmDNAPhysics}, which contrasts with the results from condensed history simulations of 18\,MeV electrons (Fig.\,\ref{fig:letelectrons} center). The resulting low-energy electron distributions originating from the spectra of the beamline simulation and the corresponding electron energy spectra in the center of the chemistry simulation box are shown in Fig.\,\ref{fig:letelectrons} bottom.\\
If present, all additional molecules such as \ce{O2} and \ce{N_2O} were simulated by background processes with the \textit{ExponentialSingleFactor} model.
All conditions were repeatedly simulated for at least thirty times with different random seeds. 
Results and values shown, are the averages from these simulations, while errorbars represent the standard error.\\
\begin{table}[!htbp]
\centering
\caption{Overview of the chemical reactions and rate constants used in the simulation of the chemical stage. The initial concentrations of the species are exemplary values, which were varied as described in the methods section. For details see the text. M denotes mol/L and \ce{H2O} is partly omitted. Values are compiled from the literature: \cite{buxtoncritical1988,vonsonntagchemical1987,planteconsiderations2017}}
\label{tab:chemistry}
 \begin{tabular}{rlccc} 
 \toprule 
 No & Reaction \& Product & Reaction rate   & Initial Concentration & Scavenging capacity  \\ 
 - & -& M\textsuperscript{-1}s\textsuperscript{-1} &  M &  s\textsuperscript{-1}  \\ 
 \midrule 
0 & \ce{e_{aq}^{-} + e_{aq}^{-}  ->   2^{-}OH + H2 }  &   1.10$\times 10^{10 }$  &  -  &  -  \\  
1 & \ce{e_{aq}^{-} + H^{.}  ->   ^{-}OH + H2 }  &   2.5$\times 10^{10 }$  &  -  &  -  \\  
2 & \ce{e_{aq}^{-} + ^{.}O^-  ->   2^{-}OH }  &   2.2$\times 10^{10 }$  &  -  &  -  \\  
3 & \ce{e_{aq}^{-} + ^{.}OH  ->   ^{-}OH }  &   3.0$\times 10^{10 }$  &  -  &  -  \\  
4 & \ce{e_{aq}^{-} + H^{+}  ->   H^{.} }  &   2.3$\times 10^{10 }$  &  -  &  -  \\  
5 & \ce{e_{aq}^{-} + H2O2  ->   ^{-}OH + ^{.}OH }  &   1.1$\times 10^{10 }$  &  -  &  -  \\  
6 & \ce{e_{aq}^{-} + HO2  ->   2^{-}OH + ^{.}OH }  &   3.5$\times 10^{9 }$  &  -  &  -  \\  
7 & \ce{e_{aq}^{-} + O2  ->   O2^- }  &   1.74$\times 10^{10 }$  &   0-270$\times 10^{-6 }$  &  0-4.7$\times 10^{6}$  \\  
8 & \ce{e_{aq}^{-} + O2^-  ->   - }  &   1.4$\times 10^{10 }$  &  -  &  -  \\  
9 & \ce{H^{.} + H^{.}  ->   H2 }  &   1.55$\times 10^{10}$  &  -  &  -  \\  
10 & \ce{H^{.} + ^{.}OH  ->   - }  &   2.0$\times 10^{10}$  &  -  &  -  \\  
11 & \ce{H^{.} + ^{-}OH  ->   e_{aq}^{-} }  &   2.2$\times 10^{7 }$  &  -  &  -  \\  
12 & \ce{H^{.} + H2O2  ->   ^{.}OH }  &   9.0$\times 10^{7}$  &  -  &  -  \\  
13 & \ce{H^{.} + O2  ->   HO2 }  &   2.1$\times 10^{10}$  &   0-270$\times 10^{-6 }$  &  0-5.7$\times 10^{6}$   \\  
14 & \ce{H^{.} + O2^-  ->   H2O2 }  &   1$\times 10^{10}$  &  -  &  -  \\  
15 & \ce{^{.}OH + ^{.}OH  ->   H2O2 }  &   1.1$\times 10^{10 }$  &  -  &  -  \\  
16 & \ce{^{.}OH + ^{.}O^-  ->   O2^- }  &   1$\times 10^{9 }$  &  -  &  -  \\  
17 & \ce{^{.}OH + H2  ->   H^{.} }  &   4.2$\times 10^{7 }$  &  -  &  -  \\  
18 & \ce{^{.}OH + ^{-}OH  ->   ^{.}O^- }  &   1.3$\times 10^{10 }$  &  -  &  -  \\  
19 & \ce{^{.}OH + H2O2  ->   H^{+} + O2^- }  &   2.7$\times 10^{7 }$  &  -  &  -  \\  
20 & \ce{^{.}OH + HO2^-  ->   O2^- + H^{+} }  &   7.5$\times 10^{9  }$  &  -  &  -  \\  
21 & \ce{^{.}OH + HO2  ->   O2 }  &   6$\times 10^{9 }$  &  -  &  -  \\  
22 & \ce{^{.}OH + O2^-  ->   ^{-}OH + O2 }  &   8$\times 10^{9 }$  &  -  &  -  \\  
23 & \ce{^{.}O^- + ^{.}O^-  ->   H2O2 + 2^{-}OH}  &   1$\times 10^{8 }$  &  -  &  -  \\  
24 & \ce{^{.}O^- + H2  ->   H^{.} + ^{-}OH }  &   8$\times 10^{7 }$  &  -  &  -  \\  
25 & \ce{^{.}O^- + H2O2  ->   O2^- }  &   3$\times 10^{8 }$  &  -  &  -  \\  
26 & \ce{^{.}O^- + HO2^-  ->   O2^- + ^{-}OH }  &   4$\times 10^{8 }$  &  -  &  -  \\  
27 & \ce{^{.}O^- + O2  ->   O3^- }  &   3.6$\times 10^{9 }$  &   0-270$\times 10^{-6 }$  &  0-9.7$\times 10^{5}$   \\  
28 & \ce{^{.}O^- + O2^-  ->   2^{-}OH + O2 }  &   6$\times 10^{8 }$  &  -  &  -  \\  
29 & \ce{H^{+} + O2^-  ->   HO2 }  &   4.78$\times 10^{10 }$  &  -  &  -  \\  
30 & \ce{e_{aq}^{-} + N_2O  ->   ^{.}OH + ^{-}OH }  &   9.1$\times 10^{9 }$  & 0-2.2$\times 10^{-2 }$  &  0-2.0$\times 10^{8}$\\  
\bottomrule 
\end{tabular} 
\end{table}
Generally, four different sets of simulations were performed to benchmark and compare the chemistry results with literature values (\textit{Set 1}), predict the DR dependent hydrogen peroxide yields (\textit{Set 2}), test different models for simulating the bunch trains (\textit{Set 3}), study g-values under LDR and UHDR conditions (\textit{Set 4}), and predict DNA radiation damage (\textit{Set 5}). 
For \textit{Set 1} simulations were performed with electrons randomly and uniformly distributed in time over 1\,$\mu$s until a target dose of 1\,mGy within the target volume was reached. The resulting DR of 1\,Gy/s represents LDR conditions.
These simulations performed in the presence and absence of oxygen (0\,mM and 270\,mM) and \ce{N_2O} as electron scavenger, to test the capability of the simulational setup to reproduce representative g-values from the literature \cite{vonsonntagchemical1987,buxtoncritical1988} for low LET and continuous LDR conditions in standard systems of pure water and with added scavengers (Tab.\,\ref{tab:gvalues}).\\
Besides the g-values of LDR conditions an important benchmark for DR dependent radiation chemistry simulations is the prediction of the hydrogen peroxide yield, which is a common endpoint due to it's long lifetime and relative accessibility by fluorescence methods involving \textit{Amplex red} or similar compounds.\cite{zhanganalysis2024} 
Experimentally it was shown in recent studies by various authors, that under UHDR the \ce{H2O2} yield is lower than for LDR, while so far all recent simulational studies showed the opposite, as discussed in detail by Zhang \textit{et al.} and the references therein.\cite{zhanganalysis2024,thomasproton2024}
These recent experiments contrast with older radiation chemical studies by pulse radiolysis some decades ago, as discussed in detail by Wardman.\cite{wardmanmechanisms2023,allenperoxide1955,wardmanradiationchemical2024}  
This points towards the urgent need for a careful evaluation of these discrepancies and experimental methodologies involved.
Therefore \textit{Set 2} aims to predict the \ce{H2O2} yield for varying DR between 0.01\,Gy/s - 10\textsuperscript{9}\,Gy/s, by simulating a uniform bunch with a bunch doses of 0.01\,Gy, 0.1\,Gy (n=300) and 1\,Gy with FWHM between 10\,ps to 1\,s, resulting in DRs between values of 1\,Gy/s (n=20, due to the high computational cost) up to UHDR values of 10$^{11}$\,Gy/s. 
The total length of the simulations were 1000\,s and the g-values were compared at the end of the simulation and shown in Fig.\,\ref{fig:chemistry-simulation-test} (top right).
In \textit{Set 3} two ways of modelling LDR train structures were compared to test if they yield the same g-values.
This was necessary since the \textit{TsIRTInterPulse} scorer currently allows only to define one repetition rate, but the experimental conditions for LDR irradiations were performed with two different repetition rates for bunch (90.9\,kHz) and trains (10\,Hz) as summarized in Tab.\,\ref{tab:irradiation}, while in the UHDR case all bunches are compressed into a single train as shown in Fig.\,\ref{fig:bunchtrainstructure}.
Therefore, a single LDR train was modeled twice, once consisting out of 9 Gaussian bunches, with a frequency of 90.9\,kHz, 6.5\,ps FWHM per bunch, and a bunch dose of 0.000563\,Gy, and again as a uniform single bunch with 0.005067\,Gy, a FWHM of  88\,$\mu$s, representing the whole train. 
For both settings g-values were compared at times of 100\,ms, the time point when the following train would start.\\
In \textit{Set 4} multiple UHDR and LDR bunch trains were simulated for total doses of up to 1\,Gy and the experimental oxygen levels of 65\,$\mu$M and 270\,$\mu$M.
In the LDR case, the train structure was modelled by the uniform train setting as in \textit{Set 3}.
For the UHDR trains the bunch dose was 0.25\,Gy, the FWHM of a bunch 25\,ps and the bunches were repeated with a frequency of 4.5\,MHz.
In \textit{Set 5} multiple UHDR and LDR bunch trains were simulated for total doses of up to 1\,Gy and the experimental oxygen levels of 65\,$\mu$M (physoxic) and 270\,$\mu$M (normoxic) under the presence of plasmid DNA with 2686\,bp and a concentration of 500\,ng/$\mu$L as used in the experiments.
The reactions of radical species were simulated by the addition of background reactions taking into account the species and DNA subunits as summarized in Tab.\,\ref{tab:DNAchemistry}.
\begin{table}
\caption{Reactions of radical species with subunits of the DNA molecules compiled from.\cite{buxtoncritical1988,vonsonntagfreeradicalinduced2006}}
\label{tab:DNAchemistry}
 \begin{tabular}{llrrr} 
 \toprule 
 No & Reaction & Reaction rate   & Concentration & Scavenging capacity  \\ 
  -& -& M\textsuperscript{-1}s\textsuperscript{-1} &  M &  s\textsuperscript{-1}  \\ 
 \midrule 
D0 & \ce{^{.}OH + Desoxyribose}  &   1.8$\times 10^{9 }$  &   1552$\times 10^{-6 }$  &  2.7$\times 10^{6}$  \\  
D1 & \ce{^{.}OH + Adenine}  &   6.1$\times 10^{9 }$  &   388$\times 10^{-6 }$  &  2.3$\times 10^{6}$  \\  
D2 & \ce{^{.}OH + Thymine}  &   6.4$\times 10^{9 }$  &    388$\times 10^{-6 }$  &  2.4$\times 10^{6}$  \\  
D3 & \ce{^{.}OH + Guanine}  &   9.2$\times 10^{9 }$  &    388$\times 10^{-6 }$  &  3.5$\times 10^{6}$  \\  
D4 & \ce{^{.}OH + Cytosine}  &   6.3$\times 10^{9 }$  &     388$\times 10^{-6 }$  &  2.4$\times 10^{6}$  \\  
D5 & \ce{H^{.} + Desoxyribose}  &   2.9$\times 10^{7 }$  &    1552$\times 10^{-6 }$  &  4.5$\times 10^{4}$  \\  
D6 & \ce{H^{.} + Adenine}  &   0.1$\times 10^{9 }$  &    388$\times 10^{-6 }$  &  3.8$\times 10^{4}$  \\  
D7 & \ce{H^{.} + Thymine}  &   6.8$\times 10^{8 }$  &    388$\times 10^{-6 }$  &  2.6$\times 10^{5}$  \\  
D8 & \ce{H^{.} + Guanine}  &   5.0$\times 10^{8 }$  &    388$\times 10^{-6 }$  &  1.9$\times 10^{5}$  \\  
D9 & \ce{H^{.} + Cytosine}  &   9.2$\times 10^{7 }$  &     388$\times 10^{-6 }$  &  3.5$\times 10^{4}$  \\  
D10 & \ce{e_{aq}^{-} + Desoxyribose}  &   1.0$\times 10^{7 }$  &    1552$\times 10^{-6 }$  &  1.5$\times 10^{4}$  \\  
D11 & \ce{e_{aq}^{-} + Adenine}  &   9.0$\times 10^{9 }$  &    388$\times 10^{-6 }$  &  3.4$\times 10^{6}$  \\  
D12 & \ce{e_{aq}^{-} + Thymine}  &   18.0$\times 10^{9 }$  &    388$\times 10^{-6 }$  &  6.9$\times 10^{6}$  \\  
D13 & \ce{e_{aq}^{-} + Guanine}  &   14.0$\times 10^{9 }$  &    388$\times 10^{-6 }$  &  5.4$\times 10^{6}$  \\  
D14 & \ce{e_{aq}^{-} + Cytosine}  &  13.0$\times 10^{9 }$  &    388$\times 10^{-6 }$  &  5.0$\times 10^{6}$  \\  
     \bottomrule 
  \end{tabular} 
\end{table}
\section{Results and Discussion}
\label{sec:discussion}
\begin{figure}[!htbp]
\centering
\includegraphics[width=0.9\textwidth]{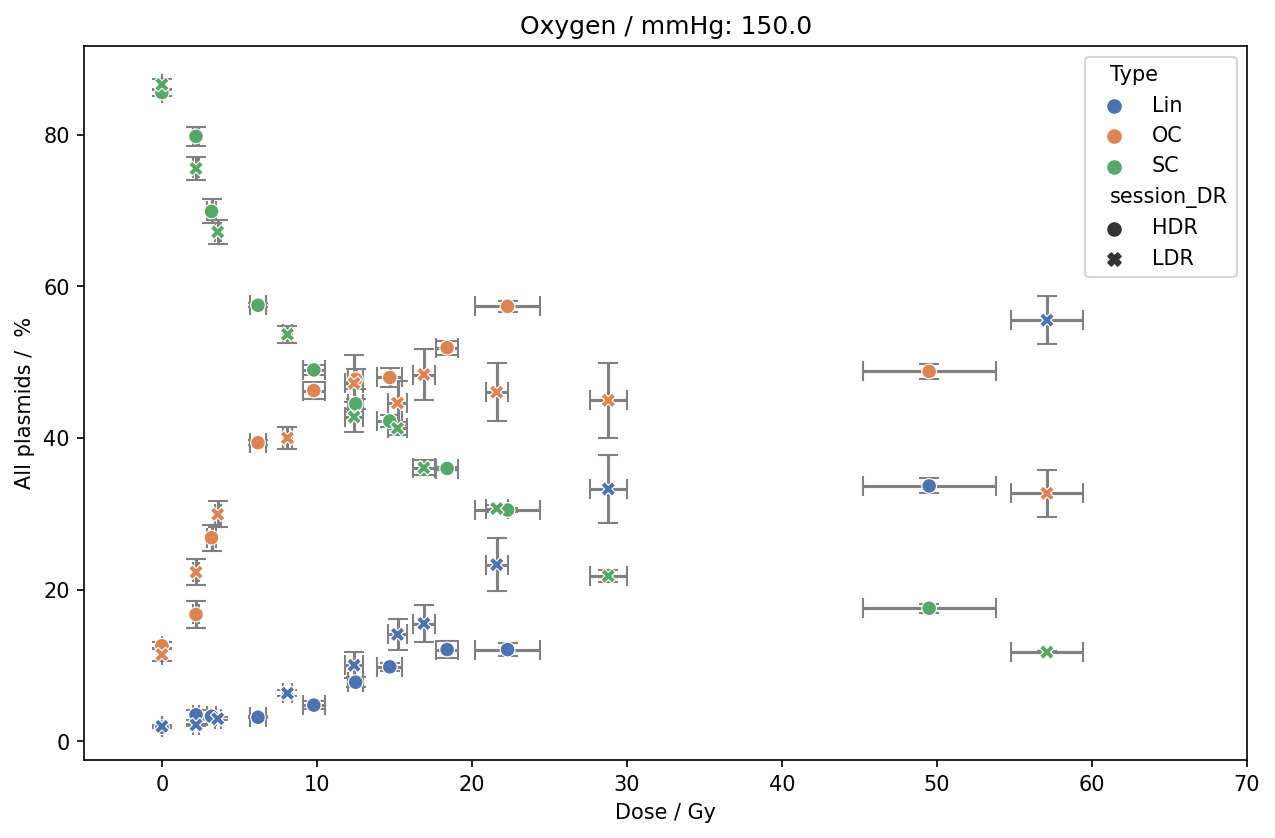}\\ 
\includegraphics[width=0.9\textwidth]{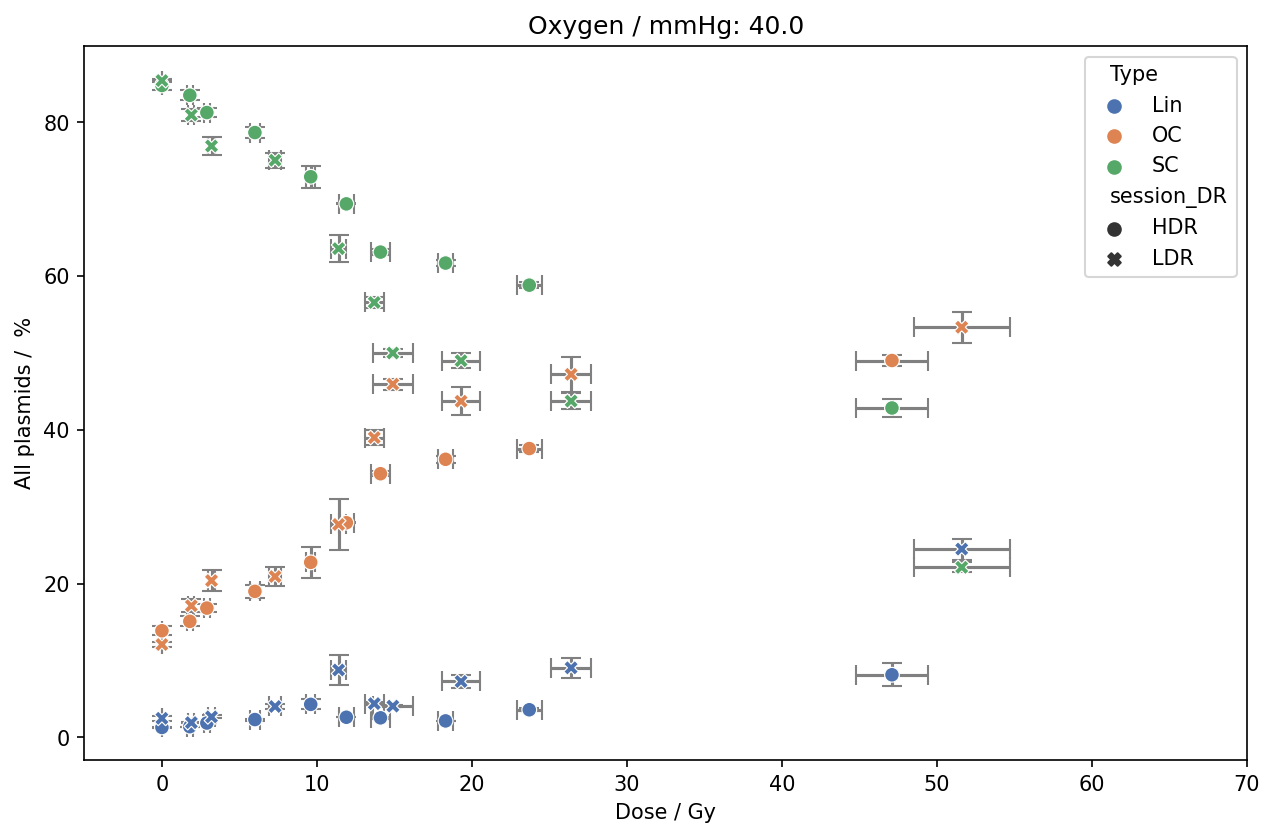}
\caption{\textbf{DNA dose-damage response:} Normalized dose damage response for plasmid DNA at ambient (top) and physiological (bottom) oxygen conditions at UHDR (circles) and LDR (crosses) shown for undamaged (SC, green), open-circular (SSB, orange), and linear (DSB, blue) DNA. Errorbars along the y-axis represent the standard error obtained from the replicates of the CGE measurements, while the errorbars at x-axis represent the measurement uncertainties obtained by dosimetry performed with Gafchromic films. For details see the text.}
\label{fig:SCOCLin-oxygen}
\end{figure}

\begin{figure}[!htbp]
\centering
\includegraphics[width=0.9\textwidth]{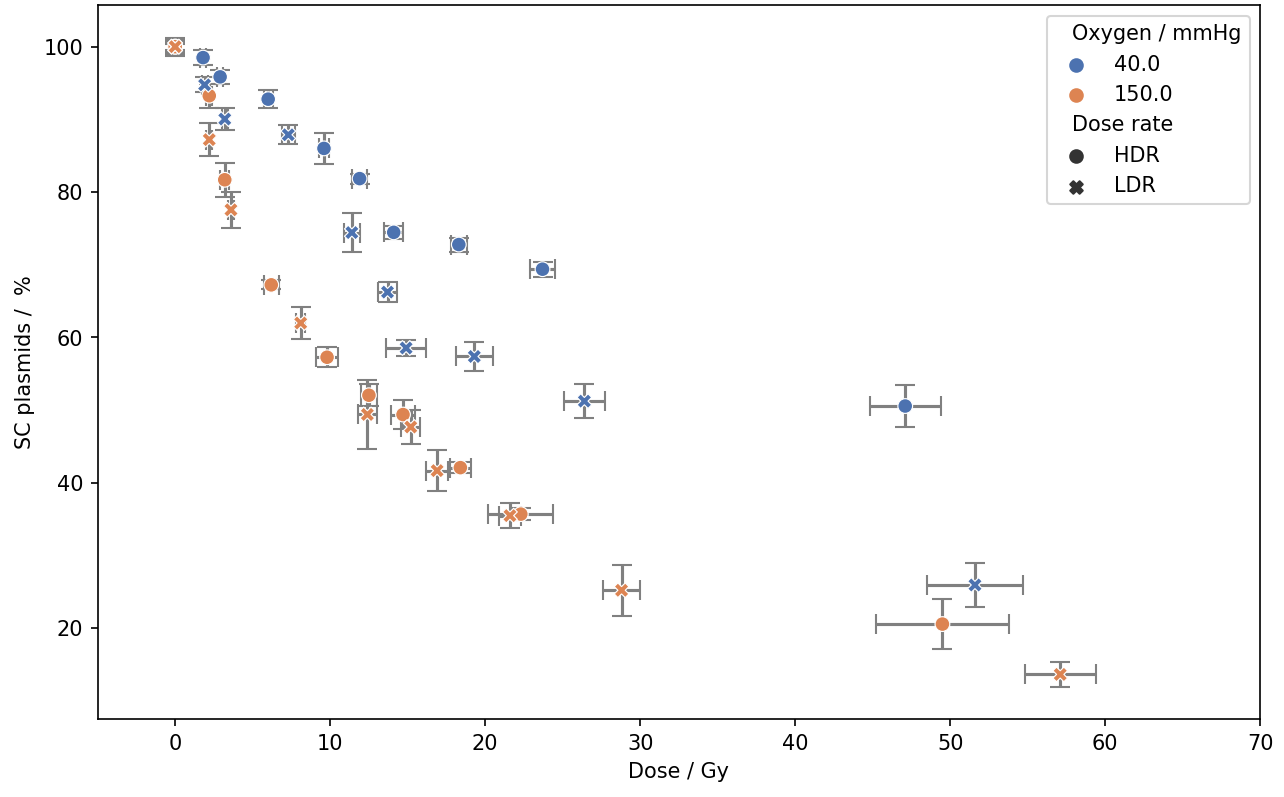}\\
\includegraphics[width=0.9\textwidth]{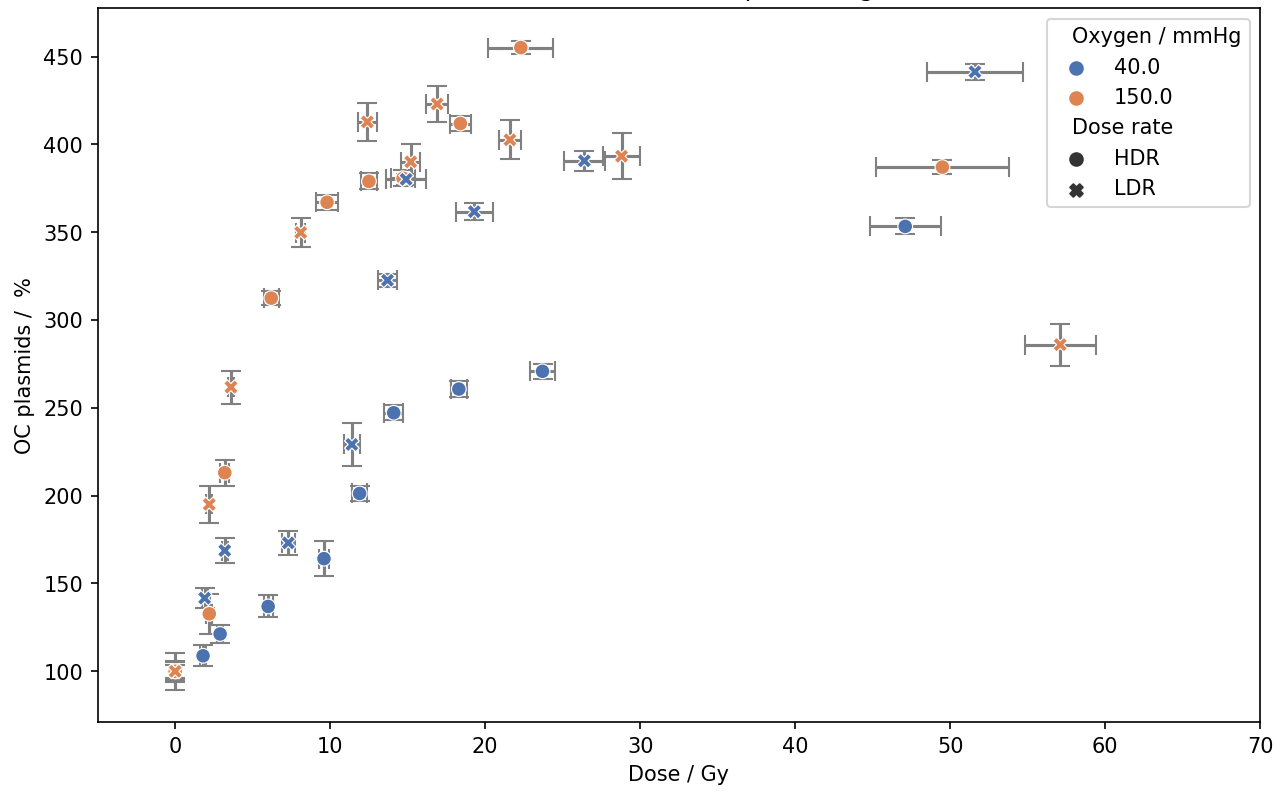}\\ 
\includegraphics[width=0.9\textwidth]{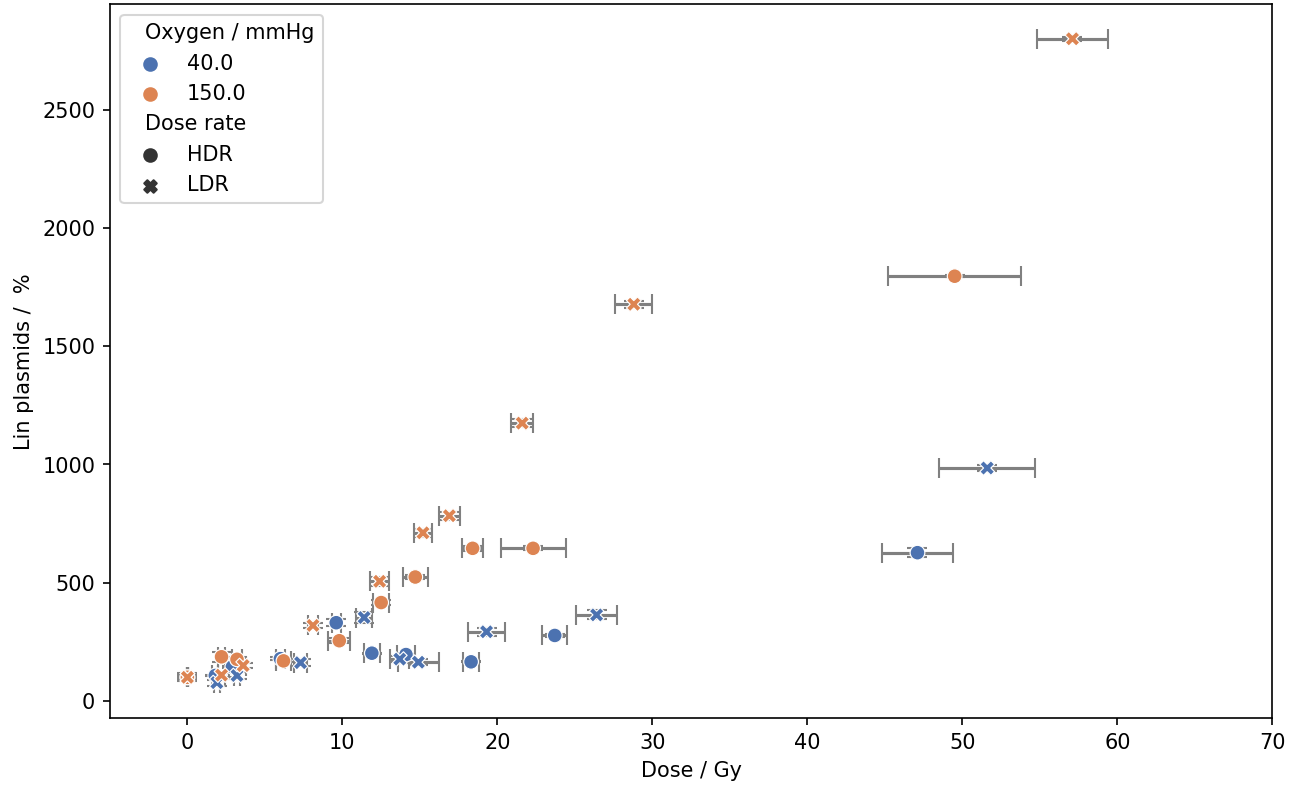}
\caption{\textbf{Relative dose-damage response:} Relative dose damage response for plasmid DNA at ambient (orange) and physiological (blue) oxygen conditions at UHDR (circles) and LDR (crosses) shown for undamaged (top), open-circular (center), and linear (bottom) DNA.}
\label{fig:SCOCLinseparated}
\end{figure}

\begin{figure}[!htbp]
\centering
\includegraphics[width=0.64\textwidth]{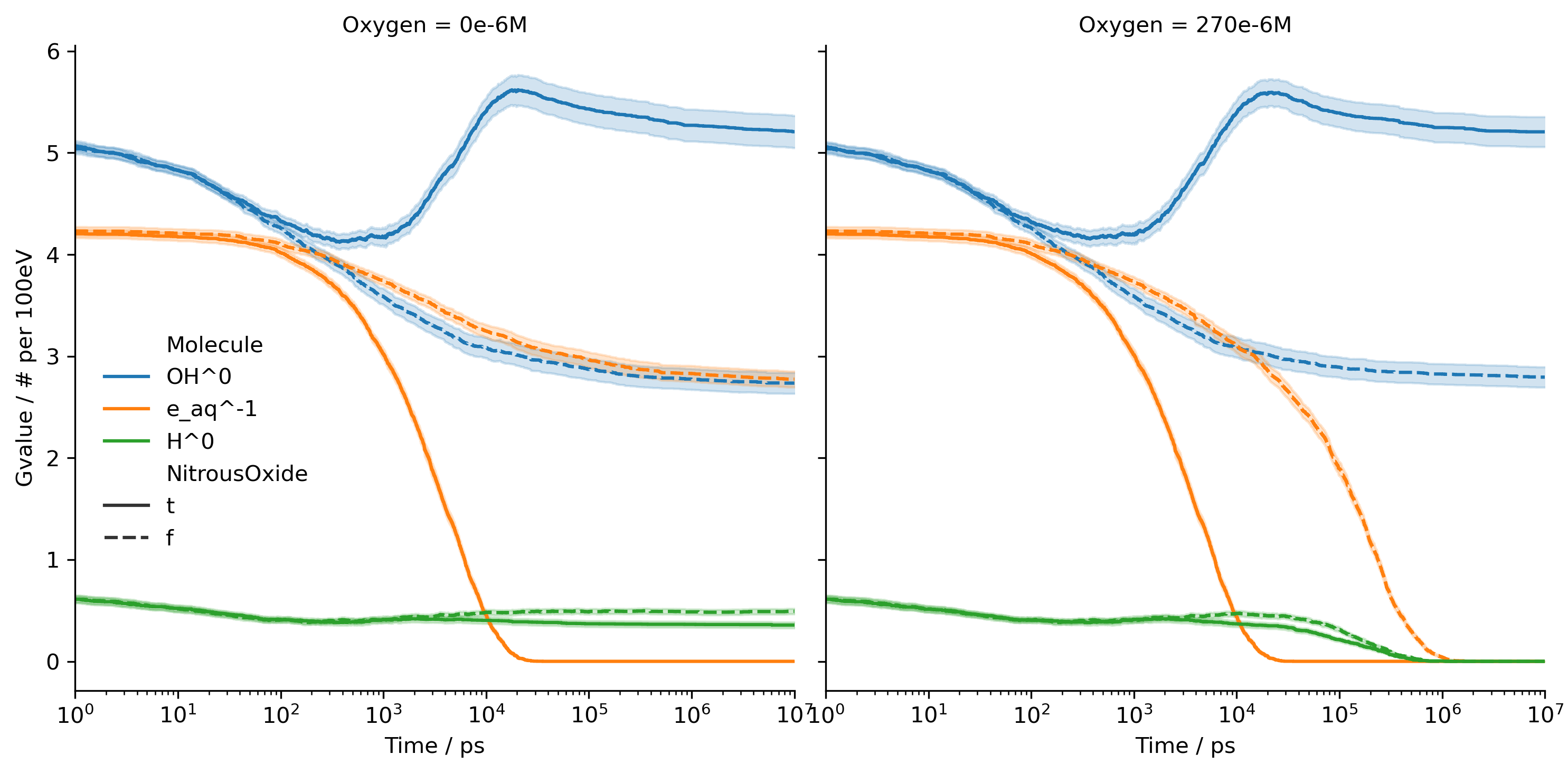}\includegraphics[width=0.35\textwidth]{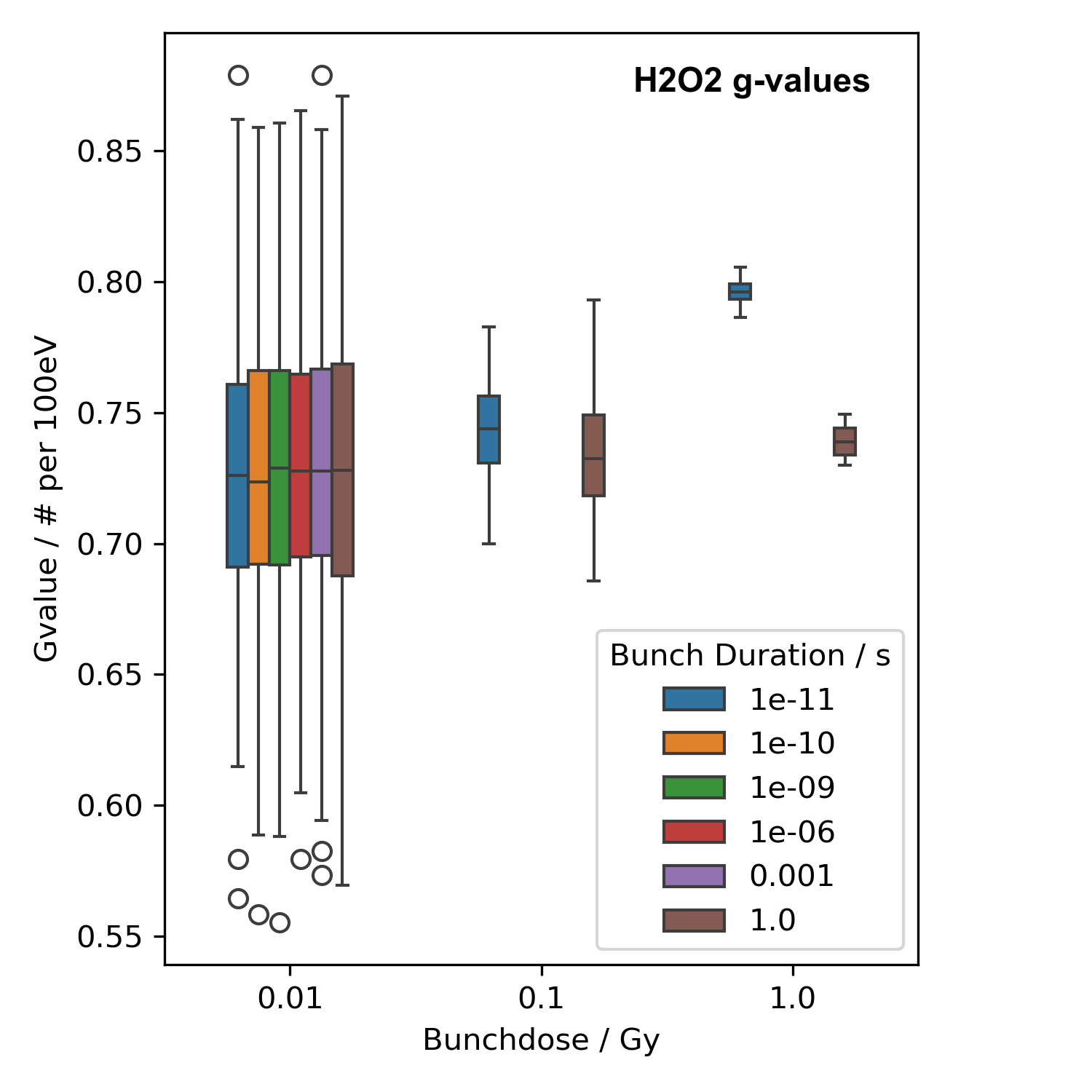}\\
\includegraphics[width=0.9\textwidth]{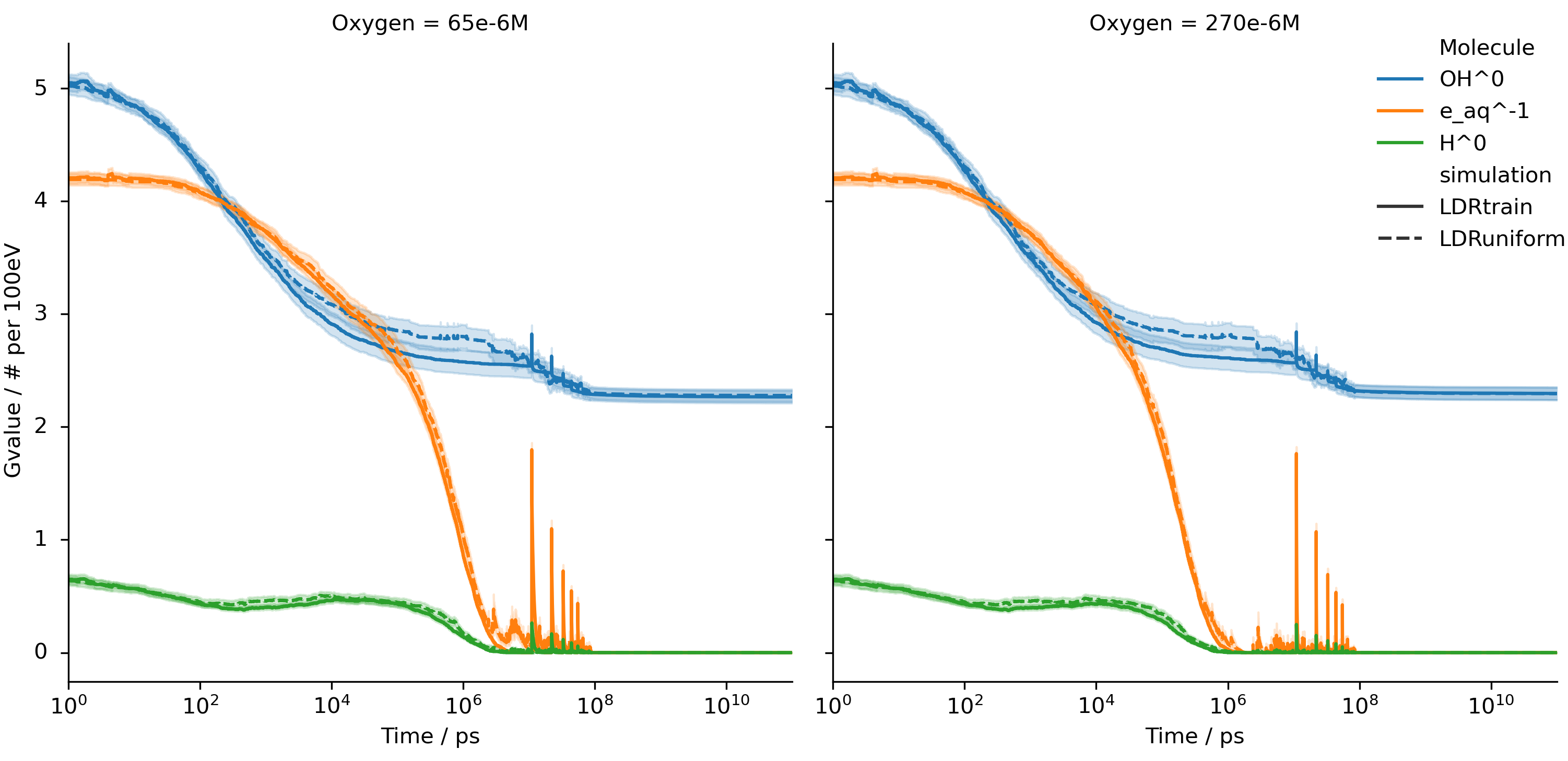}\ 
\caption{\textbf{G-values of the main reactive species} \ce{^{.}OH} (blue) \ce{e_{aq}^-} (orange) and \ce{H^{.}}(green). \textit{Set 1} (top left) shows the effect of the addition (solid line) of nitrous oxid and  oxygen (right) on radical yields under standard conditions reported in the literature. G-values at 1\,$\mu$s are reported in Tab.\,\ref{tab:gvalues}. \textit{Set 2} (top right), shows  the g-values for \ce{H2O2} at 1000\,s in dependence of bunch dose and duration (FWHM) of a uniform bunch, resulting in DRs between 1\,Gy/s (blue) up to 10$^{11}$\,Gy/s (brown). In \textit{Set 3} (bottom) the simulation of LDR by a discrete train (solid line) and uniform structure (broken line) at physiological (left) and ambient (right) oxygen condition are compared. For details see the text.}
\label{fig:chemistry-simulation-test}
\end{figure}

In the present work we performed the first irradiation and analysis of DNA under physiological salt and pH levels with LDR and UHDR electron beams at ambient and physiological oxygen conditions.
The analysis revealed a dose-rate and oxygen dependence of DNA damage in the form of strand-breaks at the DNA sugar-phosphate backbone.\\
\paragraph*{DNA irradiation in dependence of dose-rate and oxygen levels}
Under ambient oxygen atmosphere the ratio between SSB and DSB varies between DR for doses above 10-15\,Gy, showing a slight tendency for higher amount of DSBs with low DR.
While at the same time, the overall loss of undamaged/SC plasmid DNA is independent of the DR (Fig\,\ref{fig:SCOCLin-oxygen} top).
For lower, physiological oxygen, this tendency increases, showing an overall difference in strand-break induction between LDR and UHDR, with an even stronger relative amount of DSB induction under LDR compared to UHDR (Fig.\,\ref{fig:SCOCLin-oxygen} bottom).
Generally, DSB are more difficult to repair than SSB, thus they have a higher biological relevance \textit{in-vivo}. Therefore, the observed difference in DSB yields may be part of an explanation for increased survival of healthy cells under physiological conditions and UHDR.\\
If the relative dose-damage response is plotted separately for the three different states of plasmid DNA (Fig.\,\ref{fig:SCOCLinseparated}, Undamaged top, SSB center, DSB bottom) the increase of SSB and DSB induction with increasing oxygen levels can be observed. 
The loss of 50\,\% supercoiled plasmids (LD\textsubscript{50\,\%}), for LDR and UHDR at ambient oxygen is achieved around 10-15\,Gy while for physiological oxygen and LDR this happens in the region of 25-30\,Gy, and for UHDR and physiological oxygen above 45\,Gy.(Fig.\,\ref{fig:SCOCLinseparated}, top), which shows a sparing effect of the DNA between UHDR and LDR at physiological oxygen. 
The trends of dose dependent SSBs are accordingly, whereby for the ambient oxygen conditions above 30\,Gy a tendency of decreasing SSB yield (Fig.\,\ref{fig:SCOCLinseparated}, center) is observed - this indicates a transition towards DSB (Fig.\,\ref{fig:SCOCLinseparated}, bottom) at these doses, which may indicate the increasing occurrence of damage by double-hit events, or additional reactions during the late homogenious chemical stage.\\
Since the present study is performed \textit{in-vitro} on a system under physiological salt and pH, in the absence of any cosolutes and molecules which may perform enzymatical repair activity, the oberved differences between DRs and oxygen levels have to be connected to the underlying radiation chemistry.
Because it is very difficult to follow the evolution of radical species on sub-microsecond timescales experimentally, we employed radiation chemical simulations to obtain information about the their yields and reactions.\\
\paragraph*{Simulation of the radiation interaction and chemical reactions}
First, the viability of the chemical simulations to predict experimental g-values was tested by simulating classical experimental setups with low LET and LDR conditions in the absence and presence of nitrous oxide as an electron scavenger, which converts hydrated electrons into hydroxyl radicals \textit{via} reaction 30 (Tab.\,\ref{fig:chemistry-simulation-test}), and the effect of oxygen which competes for the hydrated electrons (reaction 7).
These effects are observed in Fig.\,\ref{fig:chemistry-simulation-test} top, where under hypoxic conditions (left) and in the presence of \ce{N_2O} (solid line), \ce{e_{aq}^-} (orange) is converted effectively into \ce{^{.}OH} (blue), while under physiological oxygen levels \ce{O2} (right) additional reaction channels for \ce{e_{aq}^-} and \ce{^{.}OH} are provided, leading to a modified \ce{H^{.}} yield (green).
These simulation results and literature values show a good agreement between each other for the varyiing oxygen levels and scavenging conditions (Tab.\,\ref{tab:gvalues}).\\
\begin{table}[!htbp]
\centering
\caption{Comparison of the g-values (Species/100\,eV) between simulations at 1\,$\mu$s and experimental literature values \cite{buxtoncritical1988} and von Sonntag.\cite{vonsonntagchemical1987} for different dose, bunch and chemical parameters. Hereby \ce{N2} represents deoxygenated conditions the presence of \ce{N_2O} acts as a scavenger for \ce{e_{aq}^-}. For details compare the text. Simulational values were rounded to the last digit, uncertainties given as standard error for n=50.}
\label{tab:gvalues}
\begin{tabular}{llrrcccc}
\toprule
Source & Type & DR &Condition & \ce{^{.}OH} & \ce{e_{aq}^{-}} & \ce{H2O2} & \ce{H^{.}} \\
\midrule
Buxton      \cite{buxtoncritical1988}      & Low LET    & LDR    &     -    &  2.70&  2.60& 0.70& 0.60 \\
von Sonntag \cite{vonsonntagchemical1987} & Low LET     & LDR    & \ce{N2}  &  2.70&  2.65& 0.70& 0.55 \\
von Sonntag \cite{vonsonntagchemical1987} & Low LET     & LDR    & \ce{N_2O} &  5.40&   -  & 0.85& 0.55 \\
\midrule
Simulation & DR &  Scavenger & Oxygen & \ce{^{.}OH} & \ce{e_{aq}^{-}} & \ce{H2O2} & \ce{H^{.}} \\
\midrule
LDR uniform &  0.01\,Gy/s & -        &    0\,mM&  2.8$\pm$0.1 & 2.83$\pm$0.08     &  0.62$\pm$0.04  & 0.49$\pm$0.03 \\
LDR uniform &  0.01\,Gy/s & \ce{N_2O} &  0\,mM&  5.3$\pm$0.2 & 0$\pm$0           &  0.89$\pm$0.06  & 0.36$\pm$0.03 \\
LDR uniform &  0.01\,Gy/s &    -      &    270\,mM&  2.8$\pm$0.1 & 0.04$\pm$0.01     &  0.64$\pm$0.04  & 0.001$\pm$0.001\\
LDR uniform &  0.01\,Gy/s & \ce{N_2O} &  270\,mM&  5.3$\pm$0.2 & 0$\pm$0           &  0.88$\pm$0.06  & 0.003$\pm$0.002\\
\bottomrule
\end{tabular}
\end{table}
Most of the radical species produced after water radiolysis are relatively short lived. 
An exception is hydrogen peroxide, which is stable over macroscopic timescales and can therefore be measured by fluorescence based assays \textit{e.g.} \textit{Amplex red}.
Therefore \ce{H2O2} represents an important benchmark to test simulation models in their predictive power for g-values of \ce{H2O2}.
However, so far there is a huge discrepancy between experiment and simulations.
While recent experiments conducted with protons, electrons and photons showed a decrease of \ce{H2O2} yields upon irradiation under UHDR conditions, simulation studies predict the contrary, an increase of \ce{H2O2} values with DRs, as discussed in more detail by other authors.\cite{zhanganalysis2024,sunnerbergmean2023}
With our present model, based on an extended chemistry list, including as well reactions with rate constants below 10$^9$\,M/s, the discrepancy between experiment and simulations decreased somewhat (Fig\,\ref{fig:chemistry-simulation-test} top right).
Our results show similar \ce{H2O2} g-values for low LET electrons at 1000\,s simulated by an initial uniform bunch. This initial uniform bunch was simulated with varying FWHM (10\,ps-1\,s), and with bunch doses between 0.01\,Gy-0.1\,Gy, thus resulting in DRs between 0.01\,Gy/s-10\textsuperscript{11}\,Gy/s.
This is a difference to previous studies, where higher \ce{H2O2} g-values with UHDR were predicted compared to LDR.
However it is still not yet in qualitative agreement with the recent experiments, which showed up to 20\,\% higher \ce{H2O2} yield for electron irradiated aqueous solutions at hypoxic conditions under the absence of scavengers.\cite{zhanganalysis2024}
Furthermore, we observe slightly higher values (\ce{g^{H2O2}_{UHDR}}=0.796$\pm$0.002\,per 100\,eV for an 10\,ps bunch) under UHDR conditions compared to LDR conditions (\ce{g^{H2O2}_{LDR}}0.739$\pm$0.002\,per 100\,eV for an 1\,s bunch) when the bunch dose increases to 1\,Gy for both cases (Fig.\,\ref{fig:chemistry-simulation-test} top right).
This is most likely due to the dominance of reaction 15 during the simulation, which creates a higher amount of \ce{H2O2} by interaction of hydroxyl radicals with each other for UHDR and high bunch doses.
A further increase for higher bunch doses is expected and therefore in line with the older pulse radiolysis studies.\cite{wardmanmechanisms2023,allenperoxide1955} 
However, in real-world experiments another mechanisms must be present which is currently either underestimated in the simulations, or completely absent.
For the former case, an underestimation of possible intertrack interactions, between solvated electrons and hydroxyl radicals (reaction 3) and and additional effect from the resulting \ce{^-OH} (reaction 18), depleting the \ce{^.OH} and actually lowering the \ce{H2O2} yield from reaction 3, as proposed by Zhang \textit{et al.}\cite{zhanganalysis2024}
This proposed behaviour was attributed to the higher diffusion constants of \ce{e_{aq}^-} (4.9\,$\times$10\textsuperscript{9}nm\textsuperscript{2}/s) and \ce{^-OH} (5.3\,$\times$10\textsuperscript{9}nm\textsuperscript{2}/s), compared to \ce{^.OH} (2.2$\times$10\textsuperscript{9}nm\textsuperscript{2}/s) which favor these species to react with other partners from different tracks already during the inhomogeneous chemical stage, when UHDR leads them to be on average closer to each other than under LDR conditions.
However, the question remains open, why this behaviour is not captured within the standard MCS approach?\\
Here, normally only diffusion dominated particle movements are considered.
However, the effect of radiation, especially with heavy particles such as ions, or UHDR, can lead to non-equilibrium conditions and additional, non-diffusion controlled transport of radical species,\cite{hahnmeasurements2017} or shock waves,\cite{deverathermomechanical2017,solovyovphysics2009} which may facilitate intertrack recombination.
These effects are currently completely overlooked, but may provide an explanation for the observed discrepancy, and will be studied in the future to improve upon this situation.
Additionally, another reason for the disagreement with the recent studies might be, that the newer experimental studies were performed using other containers, mostly made out of various types of plastic, than carefully cleaned glassware, which may lead to leakage of oxygen or organic species from the surface into the solvent, and in turn result in scavenging of radicals and naturally influence the related g-values as well.\cite{wardmanmechanisms2023,allenperoxide1955} 
Therefore, clearly a careful work evaluating differences between old pulse-radiolysis studies, the recent FLASH related work, and the MCS is urgently needed.
\\
As a third step the bunch train structure for the LDR irradiation conditions were modeled and compared by two approaches to test if they yield the same g-values.
This was necessary since, the implementation of the bunch scorer in the MC framework used, currently allows to define only one repetition rate, but the experimental conditions for LDR irradiations were performed by two different superimposed repetition rates for bunch (90.9\,kHz) and bunch trains (10\,Hz) (compare Tab.\,\ref{tab:irradiation} and methods section).
Therefore, a single LDR train was modeled twice, once consisting out of separate Gaussian bunches, and again as an uniform single bunch, both with the overall same dose per train and length.
The results are shown in Fig.\,\ref{fig:chemistry-simulation-test} bottom, where it is evident, that the amount of the different radical species is the same within simulational uncertainty between both approaches, especially at the time of 100\,ms, after which the subsequent train begins.
Thus, the approach to approximate the LDR bunch train structure by an uniformly distributed energy deposit over the train length is justified for the LDR conditions presented here.\\
\begin{figure}[!htbp]
\centering
\includegraphics[width=0.9\textwidth]{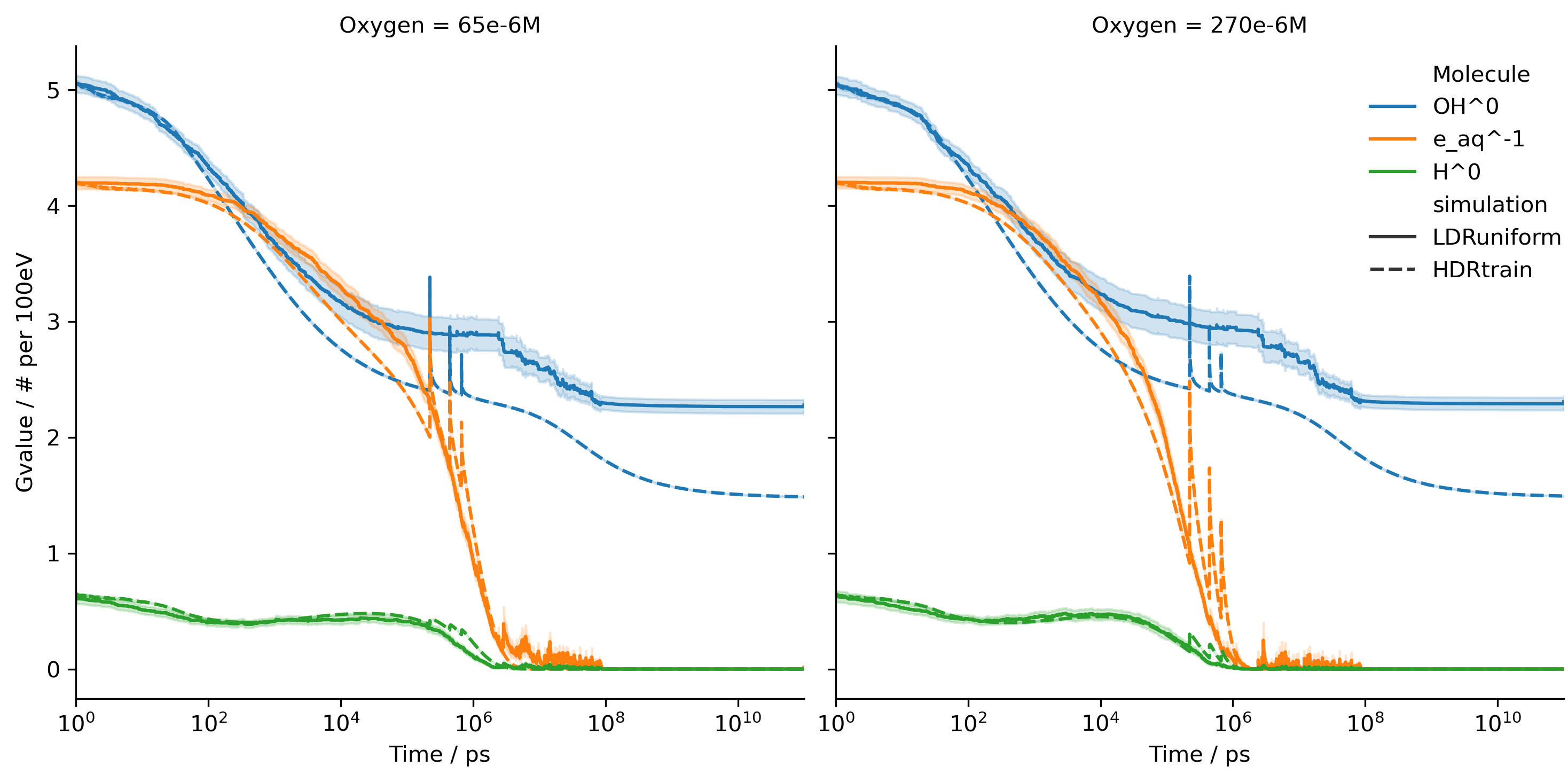}
\caption{\textbf{DR dependent g-values} of the main reactive species \ce{^{.}OH} (blue) \ce{e_{aq}^-} (orange) and \ce{H^{.}}(green) for LDR (solid line) and UHDR (broken line) at physiological (left) and ambient (right) oxygen condition. For details see the text.}
\label{fig:chemistry-simulation-PITZbeam}
\end{figure}
Based on this result, the experimentally applied conditions for LDR and UHDR irradiations were simulated for a total dose of 1\,Gy.
The evolution of the radical yield displayed in Fig.\,\ref{fig:chemistry-simulation-PITZbeam} shows a lower g-value for hydroxyl radicals (blue) for UHDR (broken line) compared to LDR (solid line) conditions, in the solution from 100\,ps onwards.
This is most likely due to an increase in the recombination of primary species under UHDR conditions as for example represented by reactions 3, 10, 12 \textit{et al.} (Tab.\,\ref{tab:chemistry}) which all have reaction rates above 10\textsuperscript{10}\,M\textsuperscript{-1}s\textsuperscript{-1}.
This interpretation is in agreement with stronger simultaneous decrease of the hydrated electrons (orange) under UHDR compared to LDR due to scavenging between \ce{^{.}OH} and \ce{e_{aq}^-} and reaction 3.
relative yield of hydrated electrons (orange) is temporarily slighlty higher for physiological oxygen conditions (left), than for ambient oxygen (right), due to the additional scavenging described by reaction 7.
This difference in reactive species, such as \ce{^{.}OH} and \ce{e_{aq}^-} may already explain the differences in SSB and DSB between the DRs since especially the hydroxyl radical is effectively causing strand-breaks in DNA.
\paragraph*{Interaction of radical species with DNA}
To understand their effects on DNA in more detail, these simulations were repeated in the presence of DNA, and the interaction of radical species which can effectively lead to a strand-break at the DNA sugar-phosphate backbone - representing the experimentally quantified endpoint in this study.
Therefore we will briefly review the most important radical species involved here, which consist under the present physiological but extracellular conditions mostly out of ROS, \ce{e_{aq}^-} and \ce{H^.}.\\
From the variety of ROS, the hydroxyl radical (\ce{^.OH}) reacts with diffusion controlled rates with the nucleobases and the desoyribose.
In contrast, superoxide (\ce{^{.}O2^-}) does not react with the DNA backbone, but can oxidize radicals already formed from other reactions at the DNA, and therefore may even protect it.\cite{cadetoxidatively2010}
Similarly to hydrogen abstraction at DNA it may oxidatively degrade lipid bilayer by oxidation and play a role in the DR dependent effects, where UHDR showed a decrease in lipid peroxidation as observed by Froidevaux \textit{et al.}
\cite{froidevauxflash2023}\\
It has to be noted here, that under the present physiological pH of 7.4 the hydroperoxyl radical \ce{HO2^{.}} is converted into the associated conjugate base, the superoxide anion \ce{O^{.-}}.
In contrast DNA lesion at the bases, \textit{e.g.} 8-hydroxyguanine and formamidopyrimidine, are generated likewise by \ce{^.OH} and \ce{^{.}O2^-}.\cite{lindahlinstability1993}
Such base damages can be converted into sugar-radicals, and can involve multiple complex and excited states as for example the conversion of an excited hole at Guanine into a base radical at the C1 position.\cite{adhikaryc52006}
Furthermore, the formation of strand-breaks by superoxide and hydrogen peroxide is possible in the presence of metal ions, \textit{via} the Fenton or Haber-Weiss reaction.\cite{leskorole1980}
However, here we used DNA in ultra-pure buffer as analyzed previously.\cite{cordsmeierdna2022}
Another possible two step process for the formation of strand-break involves an initial base loss, leading to an (abasic/apurinic) AP side, which can then be converted into a SSB by beta elimination.\cite{lindahlinstability1993}
Similarly, sugar radicals at the C4 position can lead to beta elimination as well.\cite{adhikaryc52006,giesehole2004}
Abasic sites arising from depurination can lead \textit{via} beta elimination to strand-cleavage - this happens even under physiological conditions, but with an extended half-life of the intermediates, compared to the basic conditions.\cite{gatesoverview2009}
However, to proceed quickly, this process needs either overall acidic conditions, which isn't the case for the physiological pH\,7.4 in the present experiments, but might be achieved temporally by radiolytic generated local acidic conditions which in turn accelerate SSB formation as discussed further below.
Furthermore, such UHDR induced 'acidic spikes' may modify scavenging behaviour and \ce{^{.}OH} radical yield substantially, which were predicted as well for Boron neutron capture therapy (BNCT) hadron therapy.\cite{islamsitu2018}
This may effectively modify the \ce{^{.}OH} yield, since the \ce{Cl^-} becomes under low pH an efficient hydroxyl radical scavenger ($4.3\times10^9\,$M\textsuperscript{-1}s\textsuperscript{-1}) \textit{via} the following reaction \ce{^.OH + Cl^- + H^+ -> Cl^. + H2O}.\cite{wardeffect1965}
Additionally, for physiological saline concentrations as present here, it was shown, that chloride related intermediates play an important role in cytotoxicity by producing additional radicals,\cite{saranradiation1997} however if this applies to the DNA damage yield in the present \textit{in-vitro} system cannot be stated with certainty at this point.
Furthermore, the reactivity of fully solvated electrons with desoxyribose is much lower and direct strand-break induction by \ce{e_{aq}^-} not observed so far.\cite{hahnsitu2021}\\
We note here, that the indirect damage from \ce{e_{aq}^-} has to be clearly distinguished from the direct damage caused by low-energy electrons (LEE) with kinetic energies above 0\,eV  \textit{via} formation of transient negative ions (TNI), or higher energy $\delta$-electrons, which can effectively damage DNA bases and produce SSB by dissociative electron attachment (DEA) or ionization.\cite{hahnaccessing2023}\\
However, since the probability of ionization and DEA is DR independent,\footnote{Here we note that UHDR, are 'ultra-high' in terms of radiation chemical processes, but not in the sense, that they induce measurable non-linear effects during the physical-stage, \textit{via} multiple inelastic scattering interaction at one reaction partner.} which stands in contrast to the DR dependence of the ROS and \ce{e_{aq}^-} yields (Fig.\,\ref{fig:chemistry-simulation-PITZbeam} top), we can focus in the present analysis on the indirect damage pathways, and assume the total amount of direct damage to be the same for all irradiation conditions presented here.\\
As already pointed out above, hydroxyl radicals are the principal reactive species for inducing strand-breaks by indirect effects.
Most importantly, hydrogen abstraction from the deoxyribose can lead to DNA strand scission under aerobic and anerobic conditions.\cite{pogozelskioxidative1998} 
Still, not every \ce{^{.}OH} which interacts or attaches to the backbone leads to a strand-break.\cite{vonsonntagfreeradicalinduced2006}
The approximate conversion to a strand-break after the formation of a \ce{^{.}OH}-desoxyribose (reaction D0 in Tab.\,\ref{tab:DNAchemistry}) compound can be estimated as about 24\,\%.\cite{d-kondomodeling2024a}
\begin{figure}[!htbp]
\centering
\includegraphics[width=0.9\textwidth]{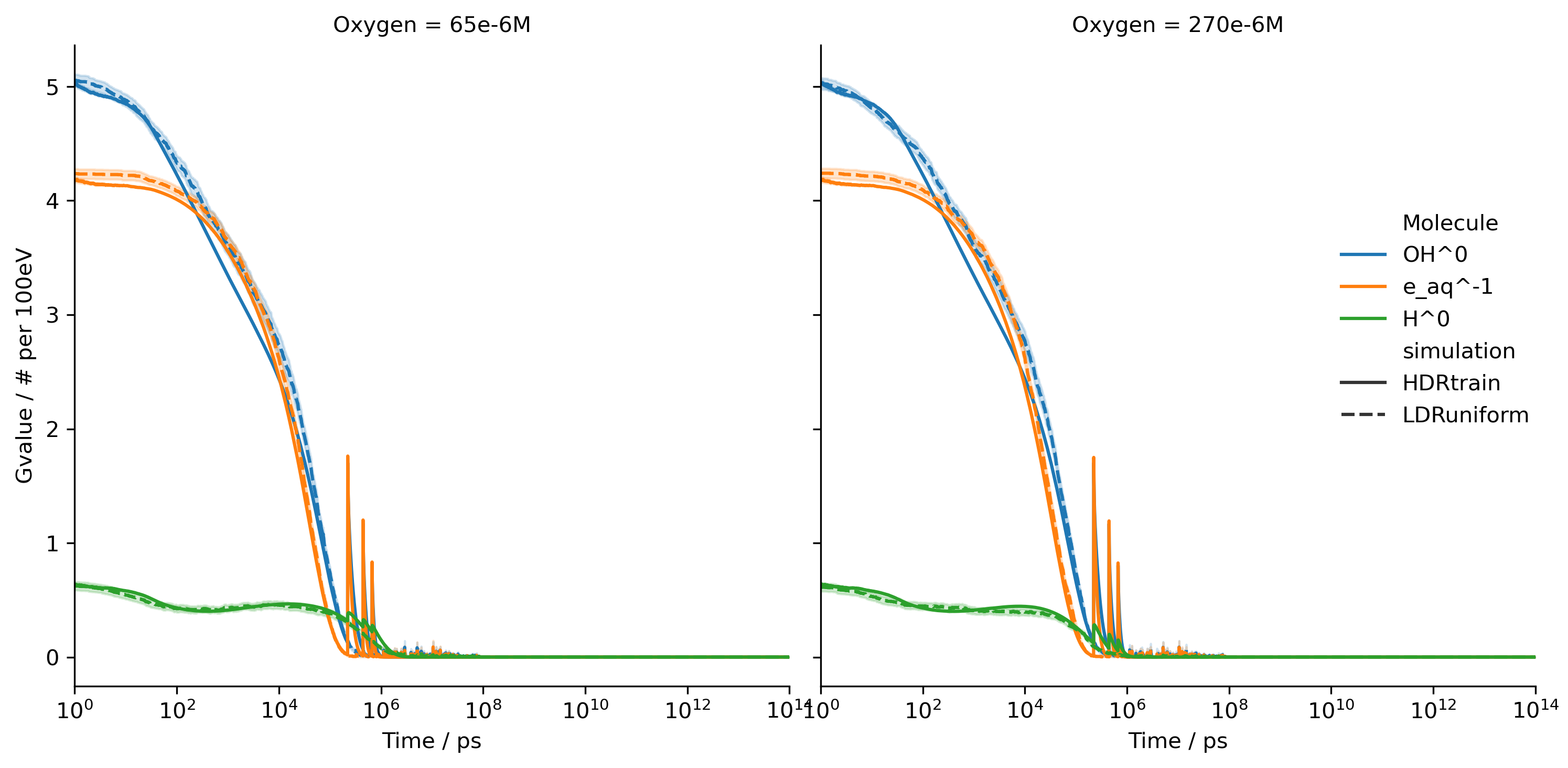}\\
\includegraphics[width=0.9\textwidth]{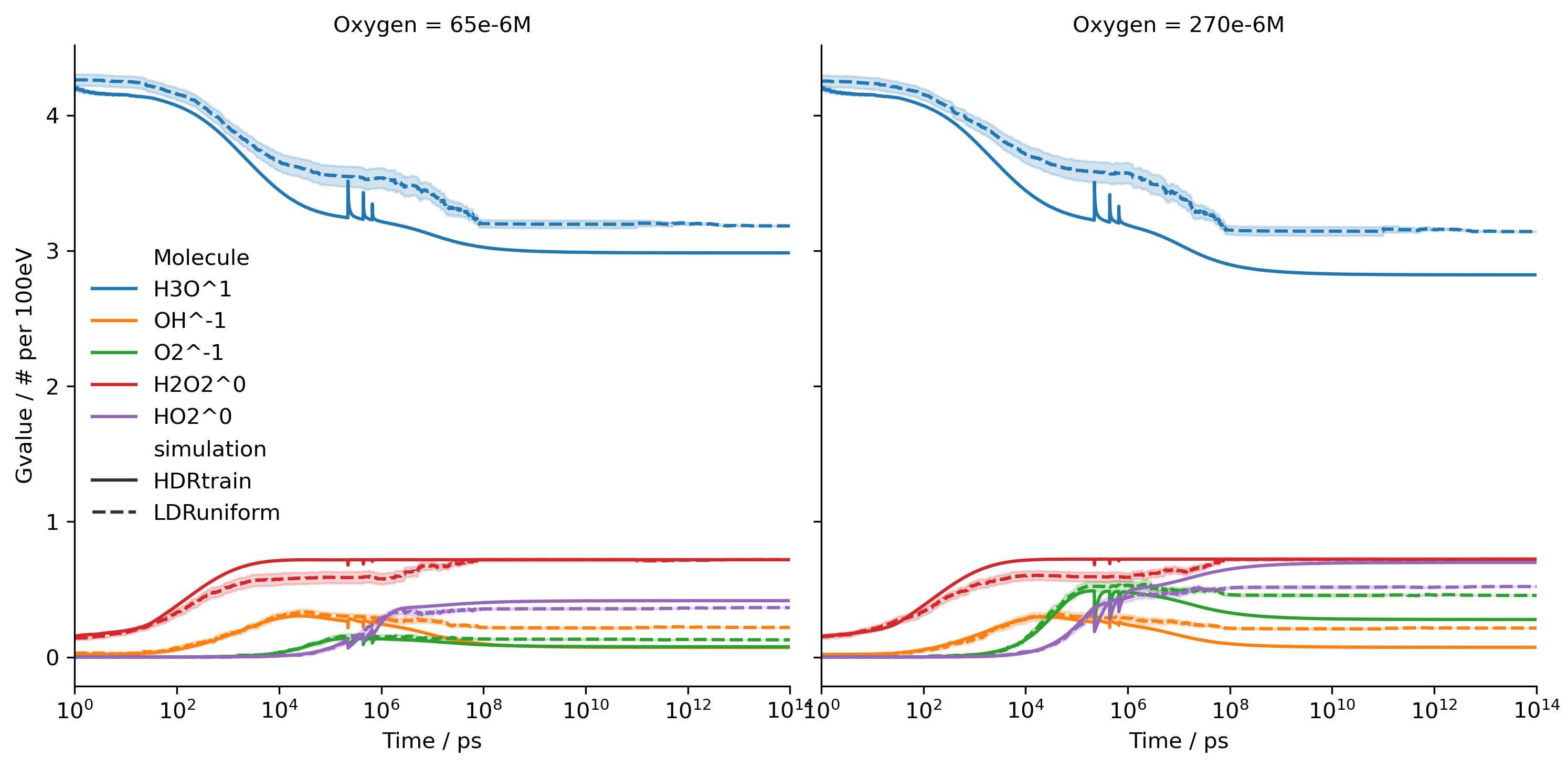}
\caption{\textbf{G-values in the presence of DNA:} G-values of the simulated reactive species as indicated in the legend, for 1\,Gy total doses under UHDR (solid line) and LDR (broken line) at physoxic (left) and ambient oxygen levels (right). Note the log-scale on the x axis. For details see the text.}
\label{fig:gvalues-with-dna}
\end{figure}
The simulated yields of ROS reactions with DNA are shown in Fig.\,\ref{fig:gvalues-with-dna}. 
Due to the hig amount of DNA in the irradiated solution (500\,ng/$\mu$L) most of the \ce{^{.}OH}, \ce{e_{aq}^-} and \ce{H^.} react with DNA within the first microsecond after generation, which contrasts with the case of DNA free solution (Fig.\,\ref{fig:chemistry-simulation-PITZbeam}).
The present conditions come closer to the crowded environment in a cell nucleus, than previous studies, which performed irradiations with an order of magnitude lower DNA concentrations,\cite{konishiinduction2023,sforzaeffect2024} which are in itself an important parameter for the overall radical yields and scavenging behaviour, in total strand break induction, as well as in terms of the SSB to DSB ratio.
This was here found to be approximately around  6:1 (at 15\,Gy where approximately 10\,\% plasmids have a DSB) under atmospheric oxygen, while a similar amount of DSB is reached only above 25\,Gy for physoxic conditions (Fig.\,\ref{fig:SCOCLin-oxygen}).
This contrasts to the SSB/DSB ratio of pUC19 DNA irradiated with 30\,keV electrons under ambient conditions which was estimated as 12:1.\cite{hahndirect2017} 
This difference may be owed to the different buffer conditions (PBS vs \ce{H2O}) and variation in time-structure (continuous emittance from a scanning-electron microscope, vs bunched structure from a linear accelerator) and LET.
\begin{figure}[!htbp]
\centering
\includegraphics[width=0.9\textwidth]{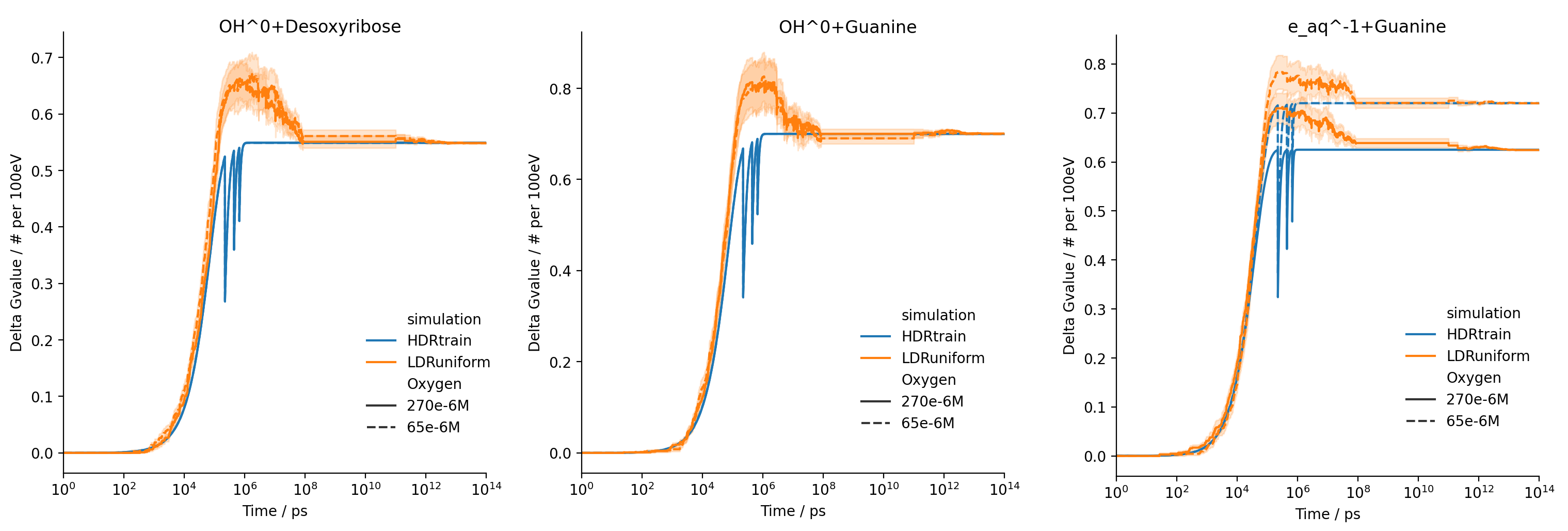}\\
\caption{\textbf{Reactions with DNA subunits:} Shown are the delta G-values (reactions between species per 100\,eV) of the hydroxyl radical (left and center) with the DNA backbone and Guanine as an exemplary nucleobase. On the right side, the reaction of the hydrated electron with Guanine is shown. LDR conditions are shown in blue, UHDR in orange, while ambient oxygen is represented with solid lines and physoxic oxygen with broken lines.}
\label{fig:deltagvalues-dna-reactions}
\end{figure}
According to the simulations, the \ce{^{.}OH} radical has a delta g-value for reaction with the backbone around 0.55\,delta g/100\,eV (Fig.\,\ref{fig:deltagvalues-dna-reactions}), which results in a g-value for DNA strand-break induction by direct \ce{^{.}OH} interaction with the backbone of around 0.13\,strand-breaks per 100\,eV deposited energy for dsDNA concentrations of 500\,ng/$\mu$L in the simulation volume of 512\,$\mu$m\textsuperscript{3}.
In contrast, the delta g-values for interaction of \ce{e_{aq}^-} and \ce{H^.} with the DNA backbone are under all conditions much lower ($<$0.015\,delta g/100\,eV, not shown) and can therefore be assumed not to contribute substantially \textit{via} this pathway.\\
As expected from the low reaction rates between \ce{H^.} and the bases, the related yield are neglible (below 0.06 for a reaction with Guanine under all conditions).
This contrasts to the behaviour of solvated electrons (Fig.\,\ref{fig:deltagvalues-dna-reactions}, right), where an oxygen dependence of the yields is observed, with a delta g-value of around 0.6 for ambient and around 0.7 for physoxic conditions for reactions with Guanine.
\paragraph*{The oxygen dependence}
Since the oxygen dependent trends for the observed strand-break induction are opposite, another oxygen dependent mechanism must be responsible for the observed effect.
Here, a two-step process is the most likely candidate: An additional pathway involves the rupture of a C-C bond in the DNA backbone by an initial hydrogen abstraction by \ce{^.OH}, followed by \ce{O2} addition to the side of the abstraction and subsequent reactions leading to an SSB.\cite{dizdarogluradiation1975,dizdaroglumechanisms2012}
In this case, the higher oxygen concentration increases the likelihood of this pathway and therefore the total conversion of \ce{^.OH} induced damage to the deoxyribose into strand-breaks beyond the aforementioned rate of 24\,\%.\\
Additionally, the potential role of singlet oxygen (\ce{^{1}O2}) in FLASH-RT, due the possible high production rate by LEE through DEA to \ce{O2}, was proposed recently.\cite{alanazisinglet2023}
Since \ce{^{1}O2} can lead to the formation of SSB and 8-hydroxydeoxyguanosin, it provides an additional possible oxygen dependent pathway for these damages.\cite{siesdamage1993}
However, at present no experimental test for this hypothesis is known.\\
Besides, the other direct reactions with the backbone (reaction D5 and D10) do not directly produce strand-breaks as already discussed above.
Due to their low yield and low damage efficiency, we can neglect them in the following as a direct cause for SSB.
Hence, for the total strand-break yield the remaining interactions between the radical species and nucleobases have to be considered.
Their products are the possible starting points for the multiple-step pathways involving the secondary ROS which may lead to additional strand-breaks.
The initial reactions contributing to base damages by chemical pathways are listened separately for Adenine, Thymine, Cytosine and Guanine in Tab.\,\ref{tab:DNAchemistry}.
We note that these values were given for isolated nucleotides and can be expected to be somewhat lower in dsDNA due to base-base interaction, effects of geometry or the presence of cations at the backbone.
These resulting secondary reactions were not included in the current simulation due to the sheer amount of possible damage types, species and pathways involved, as well as the lack of a complete set of the respective reaction rates.\cite{vonsonntagchemical1987,vonsonntagfreeradicalinduced2006}
Therefore we did not simulate them in detail and discuss the implications of these reactions, based on the general mechanisms described so far, the amount of DNA radical interactions and the yields of secondary ROS as shown in Fig.\,\ref{fig:chemistry-simulation-PITZbeam}.\\
Other secondary reactions such as superoxide interaction with neutral guanine radicals, with a rate constant of 4.7$\times$10$^8$\,M/s, leads to the formation of oxidative modifications of the respective bases and produces imidazolone and 8-oxo-7,8-di-hydroguanine.\cite{misiaszekoxidative2004}
\paragraph*{The dose-rate dependence}
The species, which showed the strongest differential response between LDR and UHDR, from the early stages on of the exposure over nanosecond timescales is \ce{H3O^+} (Fig.\,\ref{fig:gvalues-with-dna}, blue curves).
It has to be noted, that the complex buffering capacity of PBS is currently not included in the simulations.
Therefore, due to the buffered environment at pH\,7.4, the individual lifetime and the net-g-value of \ce{H3O+} is expected to be lower on microsecond timescales. Especially when the additional production of \ce{^-OH} in the radiation track is taken into account which becomes relevant at around 1\,ns (Fig.\,\ref{fig:gvalues-with-dna}, orange) 
Still, due to the high amount of base lesions, produced early on (Fig\,\ref{fig:deltagvalues-dna-reactions}), the temporary but simultaneous production of \ce{H3O^+} may lead to the additional strand-breaks by enhanced beta elimination as discussed above.\cite{lindahlinstability1993,gatesoverview2009}\\
\begin{figure}[!htbp]
\centering
\includegraphics[width=0.48\textwidth]{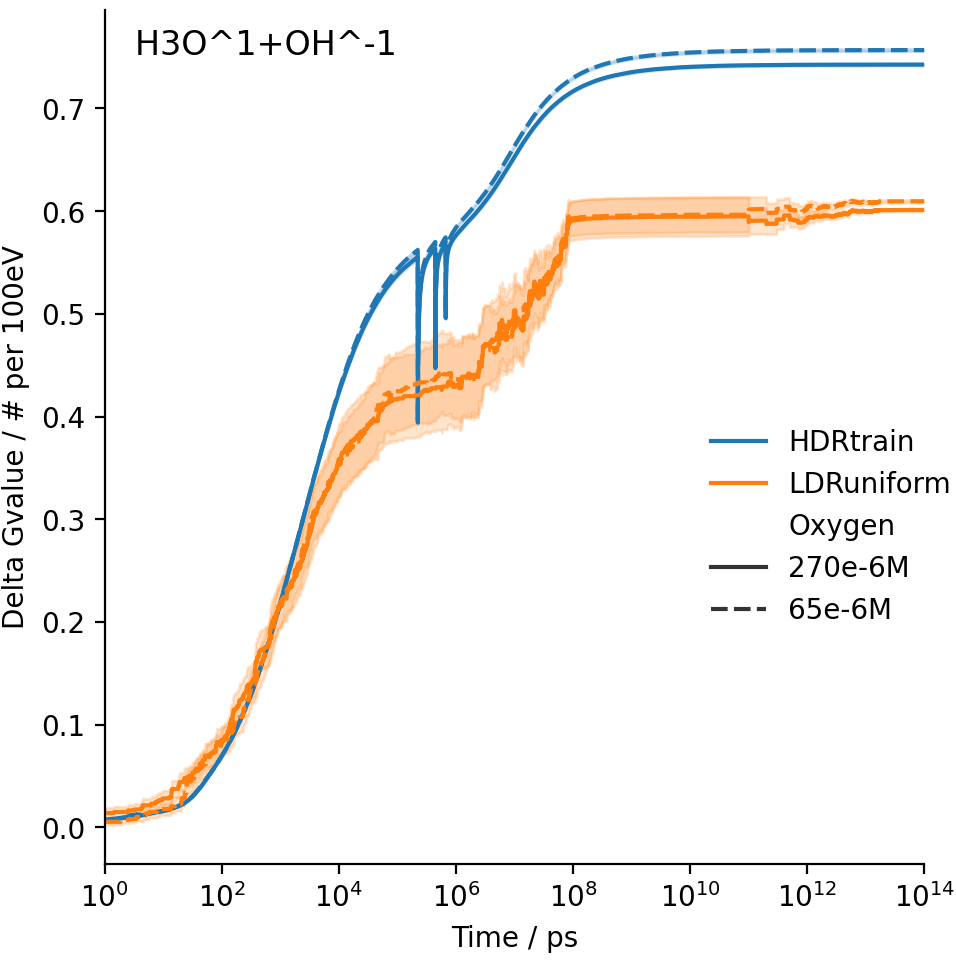}
\includegraphics[width=0.48\textwidth]{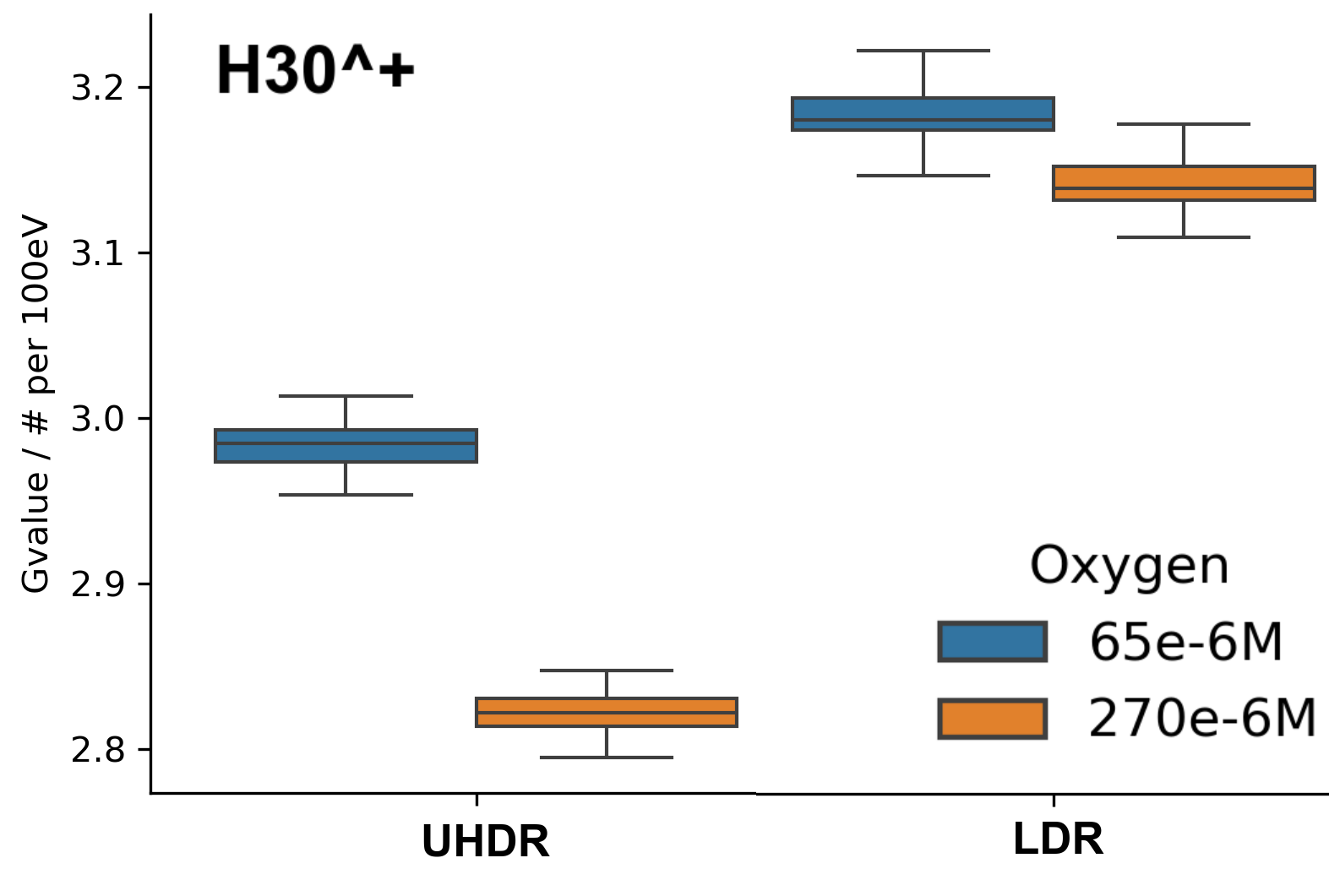}
\caption{\textbf{DR dependent \ce{H3O^+} reactions and yields.} Left: DR and \ce{O2} dependent reaction between \ce{H3O^+} and radiolytically produced \ce{^-OH} resulting in \ce{H2O}. Right: Simulated \ce{H3O^+} g-values at the end of the chemical stage for UHDR (left) and LDR (right) and ambient (orange) and physiological (orange) oxygen level as used throughout the experiments.}
\label{fig:h3oyield}
\end{figure}
Since the \ce{^.OH-desoxyribose} yields themself are DR and oxygen independent (Fig\,\ref{fig:deltagvalues-dna-reactions} left), the DR dependent production of \ce{H3O^+} (Fig.\,\ref{fig:h3oyield}) and related $\beta$ elimination induced strand-breaks are the most probable explanation for the measured DR dependence of the DNA damage.\\
\paragraph*{Irradiations under varying solvents}
How do the observed damage yields and proposed mechanisms fit into previous \textit{in-vitro} studies of UHDR effects on DNA?
Due to the overall lack of comparable studies under physiological buffer, salt and oxygen conditions we primarily compare our results to the most recent studies approaching these conditions most closely.
These have to fullfill at least two criteria which are a prerequisit to discuss the presented model in a meaningful manner: The oxygen conditions have to be well controlled and reported, and the solvent should include at least one buffering component, \textit{e.g.}\,phosphates or Tris, and stabilize the pH at physiological values.
Otherwise the pH can vary strongly even within the same environment by uptake of ambient \ce{CO2} which then leads quickly to a lowering of the pH. Thus, the likelihood of beta elimination may then increase drastically. 
Since we proposed that the sparing of DNA under UHDR is due to a lower \ce{H3O^+} production, and the following decrease in beta elimination related strand-breaks may vanish and eliminate the DR dependent effects.
Thus a controlled pH is crucial to test this hypothesis.
These assumptions are somewhat in line with the results reported by Sforza \textit{et al.} and Konishi \textit{et al.}, which studied x-ray and proton irradiated DNA in Tris buffer in the absence of salt.\cite{sforzaeffect2024,konishiinduction2023}
Konishi \textit{et al.} performed proton irradiation in a plasmid DNA system under the presence of Tris at ambient oxygen. They analyzed DNA base damage and found a sparing effect under 58\,Gy/s proton irradiation with respect to base damage and SSB, but no effect was found for the biological relevant DSB.\cite{konishiinduction2023}\\
Sforza \textit{et al.} irradiated plasmid DNA with x-rays.\cite{sforzaeffect2024}
At ambient oxygen they observed lower SSB yiels at 55\,Gy/s  compared to 0.1\,Gy/s for total doses above 60\,Gy, while at hypoxic conditions no sparing was observed.
The difference between the results under ambient conditions is most likely due to the absence of salt in their experiments.
The environment around the DNA, and especially the presence of salt is crucial for stabilizing the DNA backbone.\cite{hahndna2017}
Especially since cations accumulate around the negatively charged phosphate backbone and may additionally decrease the likelihood of \ce{H3O^+} related processes there by screening the charges. 
In contrast, their results under hypoxic conditions did not show any DR dependence.
Still, the overall strand-break yield decreased, in line with our results shown for physiologic conditions.
However, since these results, are based on complete oxygen removal by extensive bubbling with \ce{N2}, they represent a completely different hypoxia oxygen regime as found preferentially in many cancerous tissues, compared to the physiological conditions where the FLASH effect was observed, and which were studied in the present work.\\
To study in future hypoxic conditions in a similar, highly controlled oxygen environment, the sealing of the sample tubes (Fig.\,\ref{fig:Oxygensealing}) and experimental procedures have to be optimized to obtain reliable data.
These details seem to be often underreported in the literature and make a meaningful assessment and comparison of results reported difficult. 
Furthermore, it should be emphasized here to avoid misinterpretation of the present data, that the observed oxygen dependent damage in pure DNA solution cannot be explained by the mechanisms found \textit{in-vivo}, such as oxygen-fixation, which depends on the presence of cellular repair mechanisms.
Therefore, another ``purely chemical'' process, such as the one mentioned above, must be acting here to lead to the results obtained during our experiments.
Thus, in an \textit{in-vivo} system, such a mechanism would act in combination with biology dependent processes involving oxygen-fixation or others.\cite{vonsonntagfreeradicalinduced2006}
In this sense, the observations made here provide a molecular explanation based on chemical reactions, but do not exclude the possible action of other additive or multiplicative mechanisms on cellular, organ or immune system level, which may lead to tissue sparing or other modification of the DR and oxygen dependent damage yields and cellular survival.
\section{Summary and conclusion}
\label{sec:conclusion}
We have presented an in-depth description of physical and chemical parameters governing the electron irradiation experiments of a model DNA system under physiological conditions at the FLASHlab@PITZ beamline.
As an experimental endpoint DNA damage was measured in the form of DNA single and double strand-breaks.
The most important outcome is that the strand-break yields showed a dose-rate and oxygen dependency.
When comparing DNA exposed to UHDR under physiological oxygen conditions (5-6\,\% \ce{O2}) SSB and DSB yields decreased compared to ambient oxygen conditions (21\,\%).
At ambient oxygen conditions the overall damage was higher than under physiological oxygen.
Surprisingly, at physiological oxygen conditions, a sparing effect of DNA damage under UHDR was observed when compared to LDR.\\
This is the first time, that irradiations of DNA \textit{in-vitro} under UHDR showed a FLASH (sparing) effect on the molecular level for induction of DNA strand-breaks for tightly controlled physiologically relevant oxygen, salt and pH conditions.
These results stand in stark contrast to all previous studies, which were only performed in ultrapure water, or in Tris buffered solution in the absence of salt, as well as either non-physiological ambient or hypoxic conditions.
Since the FLASH effect is observed  \textit{in-vivo} as a sparing of healthy tissue, it is of utmost importance, that the oxygen, salt, and pH conditions have to resemble these conditions, as shown in this study.\\
We proposed an explanation for the observed DR and oxygen dependent behavior based on our experimental data and a combination of particle-scattering and chemistry simulations. These are summarized by the following mechanisms:
\begin{enumerate}
 \item The principal DR and oxygen independent damage is caused primarily by \ce{^.OH} radical damage to the sugar-phosphate backbone, and to a lower degree by direct effects from electrons.
 \item The observed difference in DNA damage between physoxic and normoxic  conditions is caused by a multi-step process, initialized by hydrogen abstraction through \ce{^.OH} at the backbone, which then involves the rupture of a C-C bond in the DNA, followed by \ce{O2} addition to the side of the abstraction and subsequent reactions leading to additional strand-breaks.
 \item The dose-rate dependent DNA sparing is mediated by lowering of the \ce{H3O^+} yield, which decreases the conversion rate of abasic sites to strand-breaks, due to the pH dependence of the beta elimination process.
 This process is of relevance under physoxic and normoxic conditions investigated here.
\end{enumerate}
\section{Outlook}
Future work will extend the presented methodology to lower, hypoxic oxygen conditions as found in cancerous tissue, a broader range of instantaneous, train and average dose-rates, as well as in-depth testing of the presented damage model about the underlying radiation chemical mechanisms, by introducing radical scavengers to these systems.
To be able to do this, improvements for long term oxygen control, sample throughput and optimization for more accurate simulation of the chemical interactions are needed.\\
Overall the in detail characterization of electron beam and chemical parameters presented here, lays the foundation for a series of future experiments aiming at the understanding of the mechanistic dose-rate and oxygen dependent radiation response of not only DNA, but as well DNA binding proteins, DNA-protein complexes and additional biomolecules under varying beam parameters, buffer conditions and scavenging environments, which can substantially influence the radiation response and will help to mimic cellular conditions more closely.\cite{vonsonntagfreeradicalinduced2006,hallierbiosaxs2023,wardmanapproaches2022}
\backmatter
\bmhead{Acknowledgements}
We are grateful to N.\,Aftab, S.\,Zeeshan, D.\,Dmitriiev, Z.\,Lotfi, J.\,Good, D.\,Kalantaryan, G.\,Vashchenko and M.\,Krasilnikov as members of the FLASHlab@PITZ shift crew and beamline scientists for they support during the measurements.
Discussions with D.\,Hallier, H.\,Seitz, A.\,Adhikary, C.\,Sicard-Roselli, and A.\,Solov{'}yov are acknowledged. We thank Y.\,Anis-Rafaat for proofreading of the manuscript.
\section*{Declarations}
\subsection*{Funding}
This work was funded by the Deutsche Forschungsgemeinschaft (DFG, German Research Foundation) under grant number 442240902 (HA 8528/2-2).
\subsection*{Competing interests}
The authors declare no competing interests.
\section*{Author contributions}
M.B.\,Hahn conceived the study, prepared the samples, planned the irradiation scheme, analysed the DNA samples, performed particle-scattering and chemistry simulations, developed the model, wrote the manuscript and obtained the funding. M.\,Gross and F.\,Stephan were involved in discussing and optimizing beam parameters. F.\,Riemer performed the dosimetry.  A.\,Hoffmann, C.\,Richard, and A. Oppelt operated the \textit{FLASHlab@PITZ} electron beamline during irradiations. M.B.\,Hahn and E.\,Tarakci performed the oxygen stability measurements. S.\,Aminzadeh Gohari, A.\,Grebynik, and E.\,Tarakci assisted with handling and control of the oxygen chamber during the experiments.
\subsection*{Materials, Data and Correspondence}
Correspondence and requests for data and materials should be addressed to the corresponding author M.B.H. (email: marc-benjamin.hahn@fu-berlin.de).
\begin{appendices}
\end{appendices}


\end{document}